\PassOptionsToPackage{table}{xcolor}
\documentclass[acmsmall,screen]{acmart}

\usepackage{algorithm}
\usepackage{algorithmic}
\usepackage{enumitem}

\hypersetup{
  colorlinks=true,
  linkcolor=black,
  citecolor=black,
  urlcolor=blue
}

\usepackage{graphicx}
\usepackage{subcaption}

\AtBeginDocument{%
  }

\setcopyright{none}
\renewcommand\footnotetextcopyrightpermission[1]{}

\acmJournal{TACO}
\acmVolume{37}
\acmNumber{4}
\acmArticle{111}
\acmMonth{8}

\begin{document}

\title{Mitigating the Bandwidth Wall via Data-Streaming System–Accelerator Co-Design}
\author{Qunyou Liu}
\email{qunyou.liu@epfl.ch}
\orcid{0000-0002-7410-502X}
\affiliation{%
  \institution{Embedded Systems Laboratory (ESL), École Polytechnique Fédérale de Lausanne (EPFL)}
  \streetaddress{Rte Cantonale}
  \city{Lausanne}
  \state{Vaud}
  \country{Switzerland}
  \postcode{1015}
}

\author{Marina Zapater}
\email{marina.zapater@heig-vd.ch}
\affiliation{%
  \institution{Institute of Reconfigurable \& Embedded Digital Systems (REDS), School of Engineering and Management Vaud, HES-SO University of Applied Sciences and Arts Western Switzerland}
  \city{Yverdon-les-Bains}
  \country{Switzerland}
  \postcode{1401}
}

\author{David Atienza}
\email{david.atienza@epfl.ch}
\affiliation{%
  \institution{Embedded Systems Laboratory (ESL), École Polytechnique Fédérale de Lausanne (EPFL)}
  \streetaddress{Rte Cantonale}
  \city{Lausanne}
  \state{Vaud}
  \country{Switzerland}
  \postcode{1015}
}


\begin{abstract}
Transformers have revolutionized AI in natural language processing and computer vision, but their enormous computation and memory demands pose significant challenges for hardware acceleration. In practice, end-to-end throughput is often limited by paged data movement and interconnect bandwidth, not just raw MAC count.
This work proposes a unified \textbf{system–accelerator co-design} approach to efficiently accelerate transformer inference by jointly optimizing a novel hardware matrix accelerator and its system integration with paged, streaming dataflows and explicit overlap of compute and transfer.
On the hardware side, we introduce \textbf{MatrixFlow}, a loosely‐coupled $16\times16$ systolic‐array accelerator featuring a \textbf{block‐based} matrix multiplication method that is page-aligned (4\,KB tiles), uses only a small ($\approx$20\,KB) on-chip buffer, and runs a pipelined schedule of DMA, compute, and DMA-out to fully utilize interconnect bandwidth, emphasizing standard DMA-driven streaming rather than large on-chip reuse.
On the system side, we develop \textbf{Gem5‐AcceSys}, an extension of the gem5 full system simulator allowing exploration of \textbf{standard interconnects} (PCIe) and \textbf{configurable memory hierarchies} including Direct-Memory (DM), Direct-Cache (DC), and Device-Memory (DevMem) modes with SMMU/TLB effects. Through co‐design, MatrixFlow’s novel dataflow and the Gem5‐AcceSys platform are tuned in tandem to alleviate data‐movement bottlenecks without requiring specialized CPU‐instruction‐set modifications.
We validate our approach with gem5 simulations on representative transformer models (BERT and ViT) across multiple data types and system setups. \textbf{Results} demonstrate up to \textbf{22$\times$ speed‐up} in end‐to‐end inference over a CPU‐only baseline and performance gains of \textbf{5$\times$–8$\times$} over state‐of‐the‐art loosely‐ and tightly‐coupled accelerators. Furthermore, we show that a standard PCIe-based host memory design can achieve $\sim$80\% of the performance of on‐device HBM memory. Overall, \textbf{paged streaming and pipeline overlap}, not large local SRAMs, emerge as the most effective knobs for efficient transformer inference under realistic system constraints.
\end{abstract}

\ccsdesc[500]{Computing methodologies~Simulation tools}
\ccsdesc[500]{Software and its engineering~Data flow architectures}
\ccsdesc[500]{Computer systems organization~Multicore architectures}
\ccsdesc[500]{Computer systems organization~Interconnection architectures}
\ccsdesc[500]{Hardware~Hardware accelerators}
\keywords{Computer Architecture, Design Framework, Memory Architecture, Accelerator}

\maketitle
\vspace{-0.25cm}
\section{Introduction}
\label{sec:intro}

Transformer models~\cite{Vaswani2017} have become ubiquitous in modern AI, powering breakthroughs in natural language processing (NLP) and computer vision~\cite{Devlin2019,Dosovitskiy2021}. Their success comes at the cost of massive computation and memory footprints~\cite{Ivanov2020,Karami2025}, such as billions of parameters and operations~\cite{Devlin2019,Dosovitskiy2021}, leading to significant data movement during inference~\cite{Ivanov2020}. As transformers find their way into real-time applications~\cite{Rock2022}, there is an urgent need for efficient and scalable hardware to meet these increasing computational and memory demands~\cite{Jouppi2017,Liu2026SigmaQuant}. However, designing accelerators for transformers is challenging because these models are both compute-intensive~\cite{Karami2025} and memory-intensive~\cite{Ivanov2020}, often pushing the limits of existing processing and memory architectures.

A key observation is that matrix multiplication dominates transformer workloads~\cite{Vaswani2017,Karami2025}. General matrix multiply (GEMM) operations, which underlie attention mechanisms and feed-forward network layers, account for the vast majority of transformer inference time~\cite{Karami2025}. This has made GEMM the prime target for optimization in both software and hardware. In fact, almost all high-performance deep learning accelerators feature specialized matrix engines, for example, Google's TPU uses a 256×256 systolic array~\cite{Jouppi2017,Kung1979}. Systolic arrays offer massive parallelism and high compute density perfectly suited to GEMM, and they have shown tremendous promise in accelerating ML workloads. The challenge, however, is that maximizing raw compute throughput is not enough---the surrounding system must efficiently supply data to keep the matrix units busy.

Previous studies have found that data movement, memory access, and interconnect delays can become critical limiting factors~\cite{Ivanov2020,Alian2018}. For example, a discrete accelerator connected via PCIe (a loosely-coupled device) relies on the host CPU and software for data transfer. The overhead of DMA setup, PCIe latency, and explicit cache management can be substantial, often accounting for more than 40\% of the total inference runtime in typical offload scenarios~\cite{Alian2018,Vasa2020}. On the other hand, tightly-coupled accelerators integrated on the system-on-chip (SoC) can avoid PCIe overhead by sharing memory with the host, potentially via coherent caches. This integration eases data sharing and programmability, but introduces its own challenges: hardware-managed caches incur area/power overheads and added design complexity, so many accelerators forego full coherence~\cite{Paraskevas2020}. Even with coherence, on-chip accelerators contend with shared memory bandwidth and may still be starved for data if the memory hierarchy is not designed to feed them. In short, focusing solely on the accelerator’s datapath while neglecting the interconnect and memory system can lead to unbalanced designs, for example, powerful matrix engines that idle cycles waiting for data~\cite{Shao2016,Xi2020}. This gap between accelerator microarchitecture and system architecture has become a fundamental performance hurdle.

Our work addresses this need through system-accelerator co-design for transformer inference. We jointly consider the accelerator’s internal architecture and its integration with a standard system platform to eliminate bottlenecks. In particular, we target the dominant GEMM operations in transformers with a specialized systolic-array accelerator, and we co-design the surrounding system (interconnect, memory hierarchy, and software stack) to keep the accelerator utilization high. This paper makes the following contributions:
\begin{itemize}

  \item \textbf{MatrixFlow Accelerator:} We introduce MatrixFlow, a custom $16\times16$ systolic array accelerator optimized for transformer inference.
  Unlike prior arrays that rely on large scratchpads or idealized data feeds, MatrixFlow deliberately uses \textbf{minimal on-chip buffering} and streams \textbf{page-aligned 4\,KB matrix tiles} directly from host memory. Its dataflow reformulates GEMM as a continuous stream of matrix blocks, enabling long, contiguous DMA bursts and reducing fragmentation. This co-optimized architecture sustains high utilization with a small silicon footprint by depending on a well‑orchestrated system pipeline rather than excessive local storage.

  \item \textbf{Gem5-AcceSys Full-System Simulator:} We develop Gem5-AcceSys, an extension of the gem5 full-system simulator, to allow for detailed \emph{system-level evaluation} of accelerators using standard interconnects. Gem5-AcceSys models a complete platform with support for PCIe communication, direct memory access (DMA) engines, an I/O memory management unit (SMMU) for address translation, and realistic DRAM timing models. This framework allows us to explore how different system configurations (e.g., PCIe bandwidth, memory technologies, NUMA effects) impact accelerator performance. By integrating the MatrixFlow accelerator model into Gem5-AcceSys, we can simulate end-to-end transformer inference scenarios with cycle-level fidelity, uncovering bottlenecks in data movement and interplay between the accelerator and system resources.

  \item \textbf{Co-Optimized Software Runtime:} A lightweight driver/runtime partitions matrices into page-aligned 4\,KB tiles that map cleanly to the $16{\times}16$ array, issues long contiguous DMA bursts, and double-buffers transfers to overlap PCIe/DRAM access with compute. The runtime amortizes DMA descriptors, improves SMMU/TLB locality, and avoids strided accesses, keeping the array fed in near-streaming fashion and maximizing effective bandwidth.

  \item \textbf{High Performance and Comparison:} Through full-system experiments on representative transformer models (BERT and Vision Transformer variants), we demonstrate up to 22$\times$ speedup in end-to-end inference latency compared to a CPU-only baseline. MatrixFlow’s co-designed approach also achieves superior performance against state-of-the-art alternatives: it outperforms a comparable loosely-coupled accelerator by more than 5$\times$ and a tightly coupled accelerator by more than 8$\times$ in delivered throughput. These gains stem from our balanced design: the combination of a right-sized matrix engine with an optimized interconnect and memory subsystem allows us to approach ideal accelerator utilization. Importantly, our evaluation spans different system setups (varying PCIe lane counts, memory types, etc.), and we find that the co-design approach consistently provides robust gains across these scenarios.

\end{itemize}

In summary, this work illustrates that \textbf{system-accelerator co-design}, grounded in full-system simulation, is essential for unlocking the performance and energy efficiency needed in next-generation intelligent systems.

\vspace{-0.25cm}

\section{Background and Motivation}
\label{sec:background}

\subsection{Accelerators for AI Workloads and Integration Challenges}
\label{subsec:accelerator}

Modern artificial intelligence (AI) and deep learning workloads (e.g., CNNs and Transformers) have spurred a proliferation of specialized hardware accelerators. GPUs were among the first broadly adopted AI accelerators, offering massive parallelism for general-purpose computation~\cite{Yu2024,yu2024dbfs}. Systolic arrays (SAs) have since emerged as a crucial architecture for matrix-intensive computations in deep learning, exemplified by Google’s Tensor Processing Unit (TPU)~\cite{Jouppi2017} and academic designs like Eyeriss~\cite{isscc_2016_chen_eyeriss}. These accelerators exploit data reuse and parallel data flow to achieve significantly higher performance and energy efficiency than general-purpose CPUs for matrix multiplication, the core operation in many ML models~\cite{Liu2025}. Broadly, accelerator integration falls into two categories — tightly coupled (on-chip, integrated with the CPU) versus loosely coupled (off-chip via a standard I/O interconnect) — each with its own trade-offs.

\textbf{Tightly-Coupled Accelerators (TCAs):} Tightly-coupled AI accelerators reside on the same chip as the host CPU (e.g., as co-processors or functional units sharing system memory), affording low-latency data access and often hardware-managed cache coherence. This close integration comes at the cost of extra CPU instruction and synchronization overhead, and performance can be limited by the host core’s memory bandwidth. 
For example, modern CPUs have SIMD extensions like NEON~\cite{NEON2005CortexA8}/AVX~\cite{Intel2012AVX}/SSE~\cite{AMD2018SSE,Intel1999SSE} that act as backend execution units to support vector processing, enabling data-parallel speedups for numerically intensive workloads.
From academic prototypes, TiC-SAT~\cite{Amirshahi2023} introduces a small systolic array as a tightly-coupled functional unit in the CPU pipeline, using custom ISA extensions to accelerate transformer model inference. Gemmini~\cite{Genc2021} attaches a RISC-V matrix multiplication accelerator as a RoCC co-processor, allowing in-order cores to offload DNN kernels to a tightly integrated systolic array that shares the CPU’s memory hierarchy. TCAs exemplify designs that minimize data movement, but require careful co-design to avoid stalling the CPU and to balance pipeline and memory usage.

\textbf{Loosely-Coupled Accelerators (LCAs):} LCAs are implemented as separate chips or connected modules that communicate with the CPU via standard I/O interconnects. They support highly specialized, large-scale designs with dedicated memory (e.g., GDDR or HBM), but require explicit data transfers and suffer from higher latency and potential under-utilization due to I/O bottlenecks. Classic industrial LCAs include discrete GPUs, Google’s TPU~\cite{Jouppi2017TPU}, and AWS Trainium~\cite{AWS2023Trainium}, all of which prioritize matrix throughput at the cost of communication overhead. More recently, startups like Tenstorrent~\cite{Mann2024Tenstorrent} have adopted similar principles in custom datacenter accelerators. Academic designs such as Eyeriss~\cite{isscc_2016_chen_eyeriss} and DianNao~\cite{Chen2014DianNao} demonstrate that even small-scale LCAs must carefully manage data movement across host-device boundaries. These efforts underscore the need for co-optimizing compute and communication when designing loosely-coupled AI accelerators.

Whether tightly or loosely coupled, the overarching challenge is feeding accelerators with data at high speed and coordinating with the system software, without idle cycles or bottlenecks. Modern AI workloads operate on massive models and datasets, so performance is often limited by data movement and memory access rather than raw compute throughput. This imbalance motivates a co-design approach: architects must jointly optimize the accelerator microarchitecture and its integration (memory hierarchy, interconnect, coherence mechanism) to minimize transfer overhead and latency. In practice, intermediate solutions such as coherent attachment (PCIe, CXL) are being explored to blend low latency with device autonomy. Fully understanding these trade-offs and devising balanced solutions requires full-system simulation and evaluation environments that model both the accelerator and the surrounding system (CPU, memory, I/O) in tandem. This holistic evaluation is necessary to guide the design of next-generation AI accelerators and their interfaces, ensuring that integration challenges are addressed hand-in-hand with accelerator performance goals.
\vspace{-0.25cm}
\subsection{State-of-the-Art Accelerator Simulation Frameworks}
\label{subsec:framework}
Simulating accelerators in a full-system context is vital to study these integration issues. To this end, researchers have developed several accelerator simulation frameworks, each addressing system integration to varying degrees.
Various extensions to gem5 enable accelerator modeling in a full-system context. Gem5-Aladdin couples gem5 with pre-RTL analytical models to rapidly estimate accelerator performance and energy~\cite{Shao2016}. Gem5-SALAM uses an LLVM IR approach to model custom accelerators for early-stage design evaluation~\cite{Rogers2020}. In contrast, gem5-RTL integrates actual RTL (via cosimulation) for accurate cycle information at the cost of simulation speed~\cite{Lopez2021}. Gem5-X targets many-core heterogeneous systems, supporting advanced memories and multiple accelerators to explore system-level effects (e.g., near-memory computing)~\cite{Qureshi2021}. Other frameworks, beyond gem5, full-stack simulators like SMAUG provide a deep learning workload environment on gem5 with links for accelerator models, to study end-to-end DNN performance, including data movement and scheduling~\cite{Xi2020}. Meanwhile, Accel-Sim combines GPU modeling with CPU host simulation, but lacks full-system detail~\cite{Khairy2020}.

\begin{table*}[htbp]
\vspace{-0.1cm}
\centering
\caption{Comparison of Gem5-based frameworks for hardware accelerator simulation~\cite{Liu2025AcceSys}}
\vspace{-0.1cm}
\label{tab:simulator_comparison}
\resizebox{\columnwidth}{!}{
\begin{tabular}{lcccc>{\columncolor{blue!10}}c}
\toprule
\textbf{Feature} & \textbf{Gem5-Aladdin} & \textbf{Gem5-SALAM} & \textbf{Gem5-RTL} & \textbf{Gem5-X} & \textbf{Gem5-AcceSys} \\
\midrule
\textbf{Accel.\ Design Level} & C++ & LLVM IR & RTL & C++ & C++, RTL \\
\textbf{Interconnect} & Basic buses & Basic buses & Basic buses & Basic buses & Basic buses, PCIe \\
\textbf{Accel.\ Address Translation} & Yes & No & No & No & Yes (SMMU) \\
\textbf{External Memory Simulator} & No & No & No & No & Ramulator/DRAMsys \\
\textbf{Kernel Driver Support} & No & No & No & Limited & Yes \\
\textbf{Multi-Channel DMA} & Yes & No & No & No & Yes \\
\textbf{Device-Side Memory} & No & No & No & Yes & Yes \\
\textbf{Full-System Simulation} & Yes & Bare-metal & Yes & Yes & Yes \\
\textbf{Accel.\ Process Model} & Integrated & Integrated & Integrated & Integrated & Child process (Multi-threaded) \\
\bottomrule
\end{tabular}}
\vspace{-0.1cm}
\end{table*}

These frameworks (shown in Table~\ref{tab:simulator_comparison}) have propelled research by enabling early evaluation of accelerator designs. However, \textbf{all of them exhibit notable shortcomings in modeling realistic system integration}:
\begin{itemize}
  \item \emph{Simplistic Interconnects:} Most frameworks assume idealized or basic bus interconnects (e.g., a simple memory bus or direct host memory access) and lack support for standard I/O interfaces like PCI Express~\cite{Alian2018}. As a result, they cannot accurately capture the latency, bandwidth constraints, and packet-based behavior of real peripheral links.
  \item \emph{Limited Memory Hierarchy Modeling:} Features such as Non-Uniform Memory Access (NUMA) or hybrid memory (e.g., device HBM vs. host DDR) are typically absent~\cite{Paraskevas2020,Steiner2022}. These simulators often treat all memory as monolithic, without modeling multiple memory types or the costs of crossing coherence domains and sockets.
  \item \emph{No DMA/Engine Support:} Direct Memory Access engines, which are crucial for offloading bulk data transfers from the CPU, are generally not modeled~\cite{Chi2022}. Accelerators in these simulators often magically read/write host memory, whereas real systems require programmed DMA operations or cache-coherent interfaces to move data.
  \item \emph{No Address Translation (IOMMU):} Address translation can introduce heavy overhead even for accelerators~\cite{Liu2024IAS}, yet these frameworks have not incorporated System Memory Management Units (SMMUs/IOMMUs) to handle virtual-to-physical address translation for accelerators~\cite{Paraskevas2020,Whitham2010}. Consequently, they ignore TLB miss overheads or driver-mediated pinning of pages, which can be significant in practice.
  \item \emph{Limited OS Modeling and Poor Scalability:} Many frameworks run in bare metal or user mode, lacking kernel drivers, interrupts, and realistic job scheduling, limiting their ability to model software overhead. Cycle-accurate or single-threaded models also scale poorly, making it difficult to simulate large workloads or multi-accelerator systems~\cite{Binkert2011}.
\end{itemize}

These gaps are evident when comparing feature support between frameworks. For example, Table~\ref{tab:simulator_comparison} contrasts gem5-Aladdin, SALAM, gem5-RTL, gem5-X, and Gem5-AcceSys. None of the pre-existing tools support a standard PCIe interconnect or accelerator address translation, and only gem5-X included any notion of separate device memory~\cite{Qureshi2021}, even without full IOMMU or DMA support. Kernel-driver integration was virtually non-existent before Gem5-AcceSys~\cite{Liu2025AcceSys}.
\vspace{-0.25cm}
\subsection{Motivation for a Unified Simulator–Accelerator Co-Design Approach}
\label{subsec:motivation}

The shortcomings of current frameworks motivate our unified solution: co-design the accelerator alongside the simulator. Rather than developing accelerators in isolation and then grappling with integration, we propose treating the system context as a first-class design concern.

Gem5-AcceSys extends gem5 with full-system support for accelerators. It models PCIe-based accelerator attachment, a multi-channel DMA engine, and an I/O MMU (SMMU) for realistic high-throughput data movement and address translation. The framework supports both host- and device-side memory and integrates cycle-accurate DRAM models (Ramulator, DRAMsim3, DRAMSys). It runs a full Linux OS, enabling evaluation of driver overhead, interrupts, and memory mapping. Table~\ref{tab:simulator_comparison} summarizes its key capabilities.
Gem5-AcceSys also supports flexible accelerator modeling: designers can integrate high-level C++ models or compiled RTL via a lightweight “Accelerator Wrapper” that exchanges commands with gem5 through shared memory. This enables rapid prototyping while preserving system-level accuracy.

Co-designing the accelerator with the simulator exposes architectural opportunities that isolated development would miss. Our MatrixFlow accelerator (see Section~\ref{subsec:matrixflow}) exemplifies this: it is a buffer-minimal dataflow-driven systolic array designed to stream data directly from host memory using PCIe and DMA. Unlike traditional designs that rely on large local scratchpads, MatrixFlow shifts storage and orchestration to the system, sustaining high throughput with minimal on-chip buffering.

\begin{figure}[htbp]
\vspace{-0.25cm}
  \centering
  \includegraphics[width=0.6\linewidth]{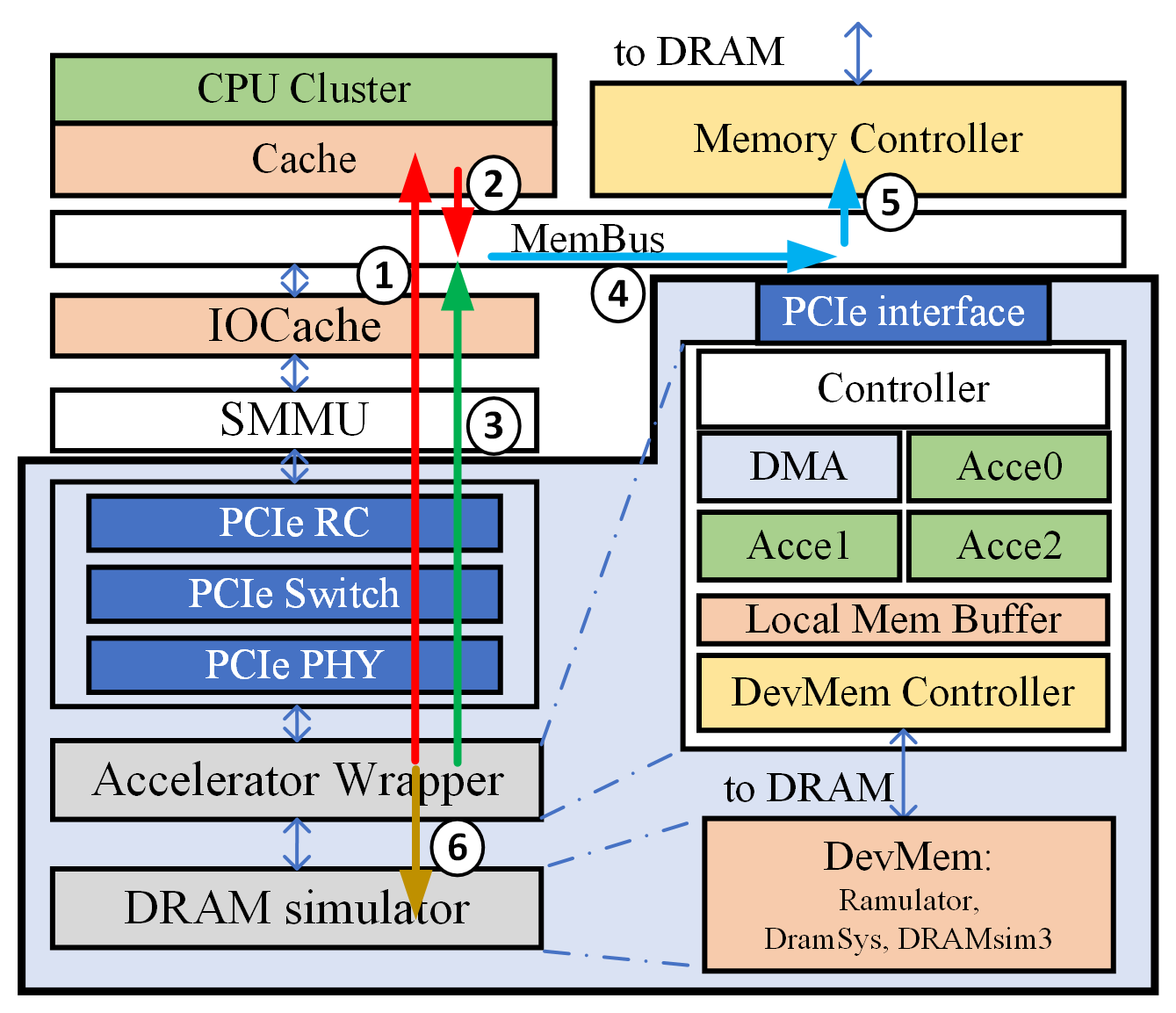}
  \vspace{-0.15cm}
  \caption{Design Framework Architecture~\cite{Liu2025AcceSys}}
  \Description{Design Framework Architecture}
  \label{fig:designFrame}
  \vspace{-0.1cm}
\end{figure}

\begin{figure}[htbp]
\vspace{-0.2cm}
  \centering
  \includegraphics[width=0.85\linewidth]{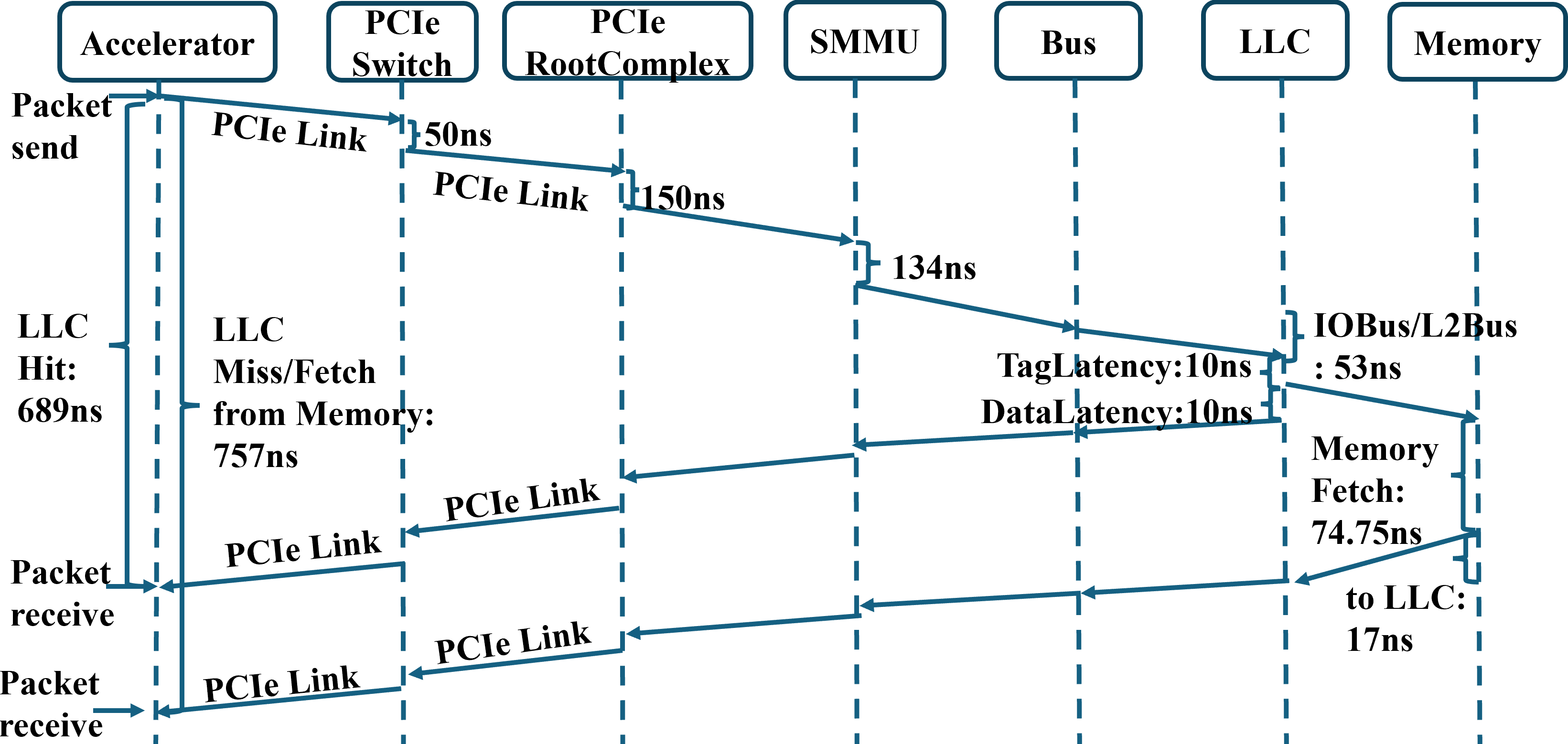}
  \vspace{-0.15cm}
  \caption{End-to-End Latency Breakdown and Message Sequence chart}
  \Description{End-to-End Latency Breakdown and Message Sequence chart}
  \label{fig:MSC}
  \vspace{-0.1cm}
\end{figure}

Figures~\ref{fig:designFrame} and \ref{fig:MSC} jointly present the system. Figure~\ref{fig:designFrame} gives the structural view and defines the datapaths (1–6).
Figure~\ref{fig:MSC} complements it with a time‑ordered message‑sequence chart for a tile, tying those datapaths to the control plane (driver doorbells, DMA descriptors, SMMU translations) and to the latency buckets used in our analysis.
We use these figures to describe the three modes—\textbf{DM} via paths (3,5), \textbf{DC} via (2,4,5), and \textbf{DevMem} via (6)—and to connect datapath activity to control events.
Section~3.2 walks through the step‑by‑step tile flow using Fig.~\ref{fig:MSC}, and §5 reports mode‑specific performance and latency breakdowns that map directly to the labeled arrows and braces in these figures. The chart shows the end‑to‑end control/data flow: accelerator data fetch $\rightarrow$ PCIe transaction $\rightarrow$ SMMU translation $\rightarrow$ LLC and the MemBus $\rightarrow$ DRAM. Braces on the right indicate the latency components. The latency buckets defined in Fig.~\ref{fig:MSC} explain where real‑system effects (PCIe bandwidth, memory latency, TLB misses) manifest in the end‑to‑end pipeline.

Evaluating MatrixFlow within Gem5-AcceSys reveals how real-world factors (PCIe bandwidth, memory latency, TLB misses) impact performance. With optimized interconnects, our loosely-coupled design can rival or exceed tightly coupled and high-memory accelerators. MatrixFlow achieves up to 400× speedup over a single-core baseline and outperforms state-of-the-art tightly (e.g., TiC-SAT) and loosely coupled accelerators. These results highlight the value of simulator–accelerator co-design. Consequently, we developed Gem5-AcceSys and MatrixFlow in tandem to ensure a balanced architecture that unifies algorithm, hardware, and system considerations. Section~\ref{sec:sysacce-codesign} details this co-designed platform.

MatrixFlow deliberately adopts a bandwidth‑driven rather than scratchpad‑driven design. Instead of large local buffers to maximize in‑place reuse, the accelerator uses a small set of page‑sized SRAMs and a page‑aligned streaming dataflow co‑designed with PCIe, multi‑channel DMA, and the SMMU. In our default implementation we provision A0/A1, B0/B1, and C—that is, 5$\times$4 KB = 20 KB—to double‑buffer the inputs while keeping one page‑sized output tile (a minimal variant uses 3$\times$4 KB without double buffering). Sizing A and B tiles to one 4 KB page lets a single DMA descriptor fetch a full tile with at most one TLB lookup, and enables a pipeline where {DMA‑in(A), DMA‑in(B), SA compute, DMA‑out(C)} overlap tile‑by‑tile.
In practice, neither the link nor the array runs at its peak continuously: the SA has fill/drain bubbles of  \(2(W-1)\) cycles per tile; the I/O path loses efficiency to descriptor/TLB overheads and PCIe packetization (e.g., non‑optimal TLP sizes), and the memory system sees bank/queueing effects.
We capture sustained effects with utilization factors \(0<\eta_{\mathrm{io}},\eta_{\mathrm{sa}}\le 1\) for the link bandwidth and the array compute, respectively. The overlap condition for a \(W\times W\) array at clock \(f\) and element size \(S\) bytes is
\begin{equation}
\frac{S\,\bigl(2WL+W^{2}\bigr)}{\eta_{\mathrm{io}}\,BW_{\mathrm{peak}}}
\;\le\;
\frac{L+2(W-1)}{\eta_{\mathrm{sa}}\,f}
\;\;\Longrightarrow\;\;
BW_{\mathrm{peak}}
\;\ge\;
\frac{S\,f\,\bigl(2WL+W^{2}\bigr)}{L+2(W-1)}\cdot\frac{\eta_{\mathrm{sa}}}{\eta_{\mathrm{io}}}\, .
\end{equation}
With \(W=16\) and \(f=1\,\text{GHz}\), the ideal asymptotes are \(\approx 32/64/128\,\text{GB/s}\) for INT8/FP16/FP32, and the effective requirement scales by \(\eta_{\mathrm{sa}}/\eta_{\mathrm{io}}\). Increasing the on-chip SRAM from \(4\,\text{KB}\) to \(8\,\text{KB}\) only raises \(L\) and tightens the ideal bound by \(\le 1\text{--}3\%\) at this design point, while doubling SRAM area/leakage; it does not remove \(\eta_{\mathrm{io}}\) losses from TLB walks, DMA descriptors, or non-optimal PCIe transactions, nor does it eliminate SA fill/drain bubbles. Consequently, we keep on-chip storage small and invest in DMA/PCIe orchestration (page-aligned tiles, tuned TLP size, double buffering) to maximize \(\eta_{\mathrm{io}}\) and \(\eta_{\mathrm{sa}}\), naturally aligned with SMMU/OS page granularity. \S3.3 details the page-blocked mapping (including the row-striped layout of \(B\)); \S5 quantifies these utilization effects with a roofline view, a 256\,B PCIe packet-size optimum, and measured TLB/translation overheads.
Under our target design point—a \(16{\times}16\) INT8 systolic array at \(1\,\text{GHz}\) with \(2\text{--}64\,\text{GB/s}\) host-memory bandwidth—using \(4\,\text{kB}\) tiles and three \(4\,\text{kB}\) on-chip SRAM buffers is sufficient to approach the roofline. Doubling the on-chip SRAM capacity to \(8\,\text{kB}\) per buffer would tighten the analytical bandwidth bound by only 1--3\% on our evaluated workloads.

\begin{table}[htbp]
\vspace{-0.15cm}
\centering
\caption{Microarchitecture positioning and memory design logic.}
\resizebox{\textwidth}{!}{
\label{tab:uarch-contrast}
\small
\begin{tabular}{p{3.0cm}p{3.5cm}p{3.4cm}p{3.6cm}}
\toprule
\textbf{Aspect} & \textbf{Gemmini (RoCC SA)} & \textbf{eGPU (SIMT)} & \textbf{MatrixFlow (ours)} \\
\midrule
Coupling to host & Tightly coupled (on-SoC, RoCC) & Tightly/loosely (SoC GPU or discrete) & Loosely coupled over PCIe \\
Compute core & Systolic array (WS/OS dataflows) & SIMT cores + tensor intrinsics & 16$\times$16 systolic array (OS) \\
Local on-chip storage & Large scratchpads + accumulator SRAMs (hundreds of kB) & Caches (I/D/L1), shared L2 & \textbf{3$\times$4\,kB} SRAM (A/B/C) \\
Locality strategy & \textbf{In-place reuse} in scratchpads; traffic shaped for SA & Cache/memory coalescing via warps & \textbf{Page-aligned streaming}; overlap DMA/compute \\
Interconnect \& translation & SoC buses; no SMMU on device & SoC interconnect or PCIe; MMU in GPU & \textbf{PCIe + multi-channel DMA + SMMU} \\
Tiling unit & Accelerator tiles sized for scratchpad & Kernel-dependent (CTA/warp tiles) & \textbf{One OS page (4\,kB)} per A/B tile; row-striped B \\
Overlap mechanism & SW/ISA-managed prefetch/reuse & Warp scheduling, async copies & \textbf{Double-buffered DMA}\{A,B\}/compute/\textbf{DMA-out} C \\
\bottomrule
\end{tabular}}
\vspace{-0.15cm}
\end{table}

MatrixFlow’s design point differs from mainstream accelerators in where locality is created and how data reaches the compute core. Table~\ref{tab:uarch-contrast} contrasts three representative points: (i) \textbf{Gemmini~\cite{Genc2021}}---a tightly coupled RoCC systolic array with large on-chip scratchpads and accumulator memories to maximize in-place reuse, support output-stationary (OS) and weights-stationary (WS); (ii) an \textbf{embedded GPU (eGPU)~\cite{machetti2025egpuopensourceconfigurableriscv}}---a SIMT core complex that relies on caches and warp scheduling for locality, trading deterministic dataflow for generality; and (iii) \textbf{MatrixFlow (ours)}---a loosely coupled, page-aligned streaming systolic array that minimizes on-chip storage and uses DMA+SMMU to overlap transfers and compute, supporting output-stationary, which has only three sets of 4\,kB local buffers (2 double-buffered SRAM and 1 normal SRAM component, with 20\,kB physical SRAM but 12\,kB logically).
This comparison showcases that the novelty of this work does not come from designing a new systolic array topology but rather from the page-aligned streaming dataflow co-designed with PCIe, multi-channel DMA and SMMU, which makes a tiny 3$\times$4\,KB SRAM sufficient while sustaining utilization once bandwidth satisfies the overlap bound.

\begin{table}[htbp]
\vspace{-0.15cm}
\centering
\caption{Local on-chip storage for data reuse (SRAM/caches)}
\label{tab:local-reuse}
\small
\resizebox{\columnwidth}{!}{
\begin{tabular}{lrrr}
\toprule
\textbf{Design} & \textbf{Local storage (kB)} & \textbf{SRAM/cache area (mm$^2$)} & \textbf{Dyn.\ power (mW)} \\
\midrule
Gemmini Scratchpad & 320 & 0.731  & 86.59 \\
eGPU caches (I\$\,4\,kB + D\$\,16\,kB) & 20 & 0.0809 & 17.70 \\
MatrixFlow buffer & \textbf{20} & \textbf{0.100} & \textbf{55.79} \\
\bottomrule
\end{tabular}}
\vspace{-0.15cm}
\end{table}

Table~\ref{tab:local-reuse} highlights the order-of-magnitude difference in local storage philosophy (hundreds of kB scratchpads vs.\ page-sized 20\,kB), while Table~\ref{tab:compute-core} shows that the compute macro in MatrixFlow remains compact and competitive in area/power; all PPA figures are synthesized with the same 28\,nm TSMC technology using our own synthesis flow. Together with the overlap bound in \S3.1, this supports a bandwidth-driven, page-aligned streaming design over scratchpad-centric reuse.

\begin{table}[htbp]
\vspace{-0.15cm}
\centering
\caption{Compute core footprint (matrix/tensor engine)}
\label{tab:compute-core}
\small
\resizebox{\columnwidth}{!}{
\begin{tabular}{lrrrrr}
\toprule
\textbf{Design (dataflow)} & \textbf{Area (mm$^2$)} & \textbf{Dyn.\ power (mW)} & \textbf{Peak (GOPS)} & \textbf{GOPS/W } & \textbf{GOPS/mm$^2$} \\
\midrule
Gemmini SA (OS/WS)        & 0.230 & 102.47 & 512.0   & 4{,}996.58 & 2{,}226.09 \\
eGPU SIMT tensor core     & 0.200 & 83.00  & 9.6   & 120 & 48 \\
\textbf{MatrixFlow SA (OS)} & \textbf{0.186} & \textbf{84.55}  & \textbf{512.0} & \textbf{6{,}050.90} & \textbf{2{,}752.69} \\
\bottomrule
\end{tabular}}
\vspace{-0.15cm}
\end{table}

\section{System-Accelerator Co-Design Architecture}
\label{sec:sysacce-codesign}
Our accelerator is a loosely‑coupled PCIe device that streams page‑aligned tiles from the host via a multi‑channel DMA and an SMMU (IOMMU) for VA$\rightarrow$PA translation. The accelerator keeps only three 4 KB SRAM tiles (see Fig.~\ref{fig:sysmy}) on chip and relies on the system interconnect + DRAM to supply data at line rate. We evaluate three access modes that correspond to the numbered arrows in Fig.~\ref{fig:designFrame}: Direct‑Memory (DM) uses path (3, 5) to go straight to DRAM; Direct‑Cache (DC) uses (2, 4, 5) to enter the LLC and fall back to DRAM on misses; DevMem uses (6) to read/write device‑side memory. This labeling scheme is used throughout \S3--\S5 to tie microarchitecture, interconnect, and software runtime together.
\subsection{MatrixFlow Accelerator Hardware Design}
\label{subsec:matrixflow}
\begin{figure}[htbp]
\centering
\includegraphics[width=0.7\linewidth]{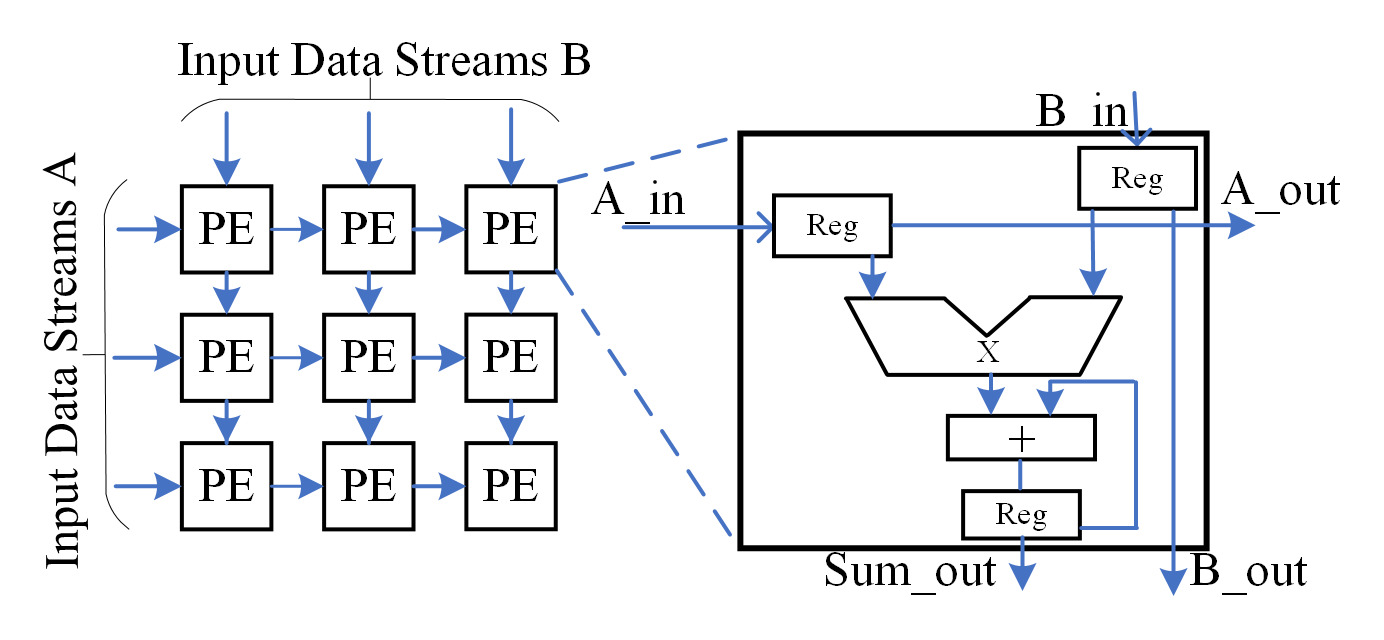}
\vspace{-0.25cm}
\caption{Systolic Array Architecture~\cite{Liu2025}}
\Description{Systolic Array Architecture}
\label{fig:systolicarray1}
\vspace{-0.25cm}
\end{figure}

MatrixFlow is a loosely coupled matrix-multiplication accelerator built around a 16×16 systolic array (SA) of Processing Elements (PEs) optimized for transformer workloads (Figure~\ref{fig:systolicarray1}).

\begin{figure}[htbp]
  \centering
  \includegraphics[width=0.4\linewidth]{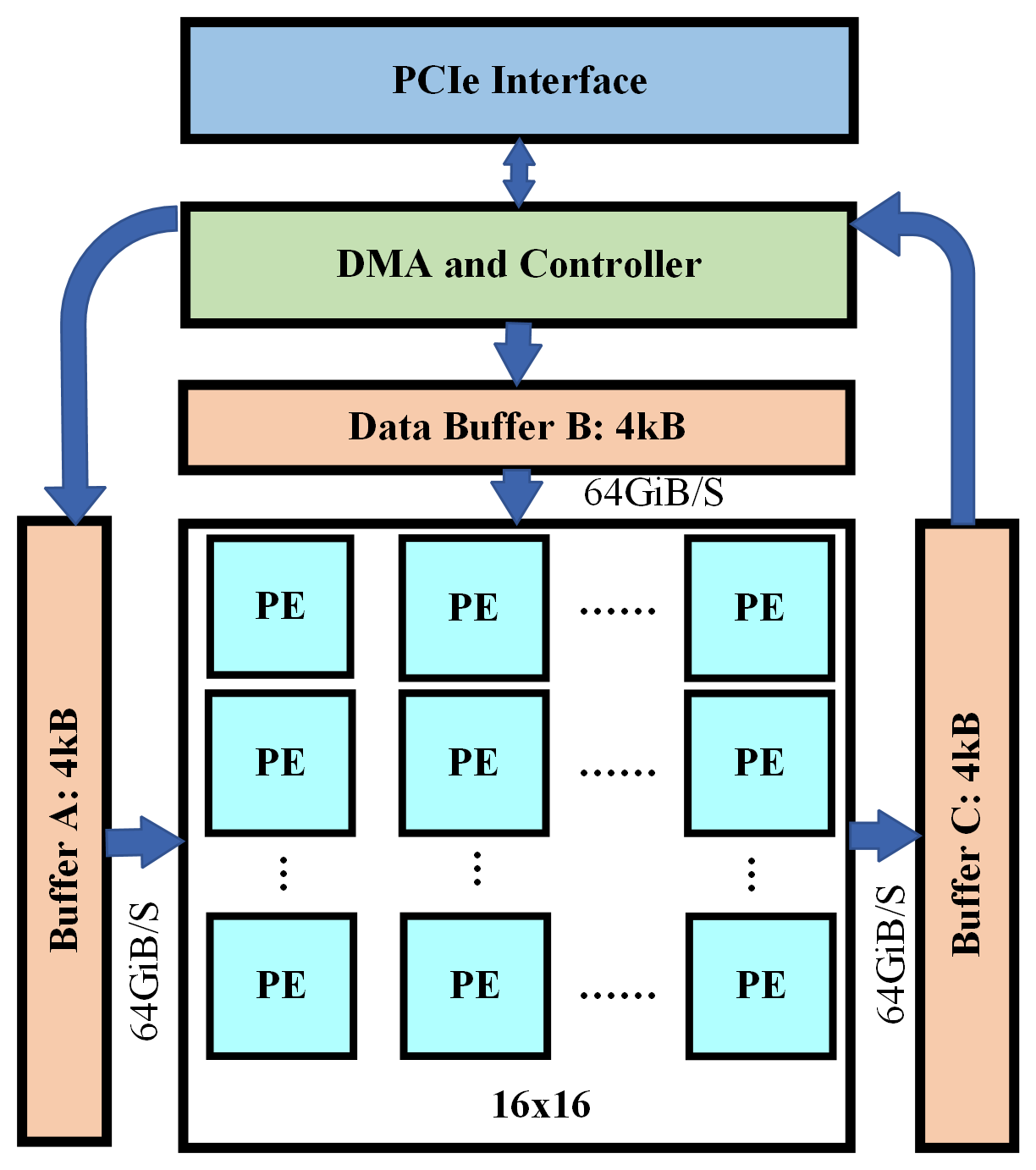}
  \vspace{-0.1cm}
  \caption{Hardware design for systolic array~\cite{Liu2025}}
  \Description{Hardware design for systolic array}
  \label{fig:sysmy}
  \vspace{-0.25cm}
\end{figure}

The accelerator is equipped with PCIe and a DMA interface logic alongside the PE array (Figure~\ref{fig:sysmy}) to support direct data streaming between host memory and the array, enabling high-throughput operation without CPU intervention.

Unlike tightly coupled accelerators requiring ISA modifications, MatrixFlow attaches to the system as a peer device, simplifying integration into full-system simulators like Gem5-AcceSys (see Section~\ref{subsec:gem5-accesys}). Notably, MatrixFlow's emphasis on efficient DMA-driven data streaming and minimal on-chip buffering echoes the design principles of Intel's Data Streaming Accelerator (DSA), which similarly couples high-throughput DMA engines with a lightweight accelerator architecture to offload data movement from the CPU.

A key novel aspect of MatrixFlow’s hardware is its minimal on-chip buffering. The accelerator includes only three small SRAM buffers (on the order of kilobytes each) for data staging, dramatically less than typical AI accelerators that often employ buffers in the hundreds of KB. Specifically, as shown in Figure~\ref{fig:sysmy}, there are two 4KB input buffers (for matrix A and matrix B tiles) and one 4KB output buffer.
Notably, 4KB corresponds to the size of a standard memory page, and by sizing the buffers to hold one page of data, the design aligns perfectly with the block-based data mapping discussed in Section~\ref{subsec:dataflow}.
The input tiles are streamed into the array, the partial results flow diagonally across the PEs, and the completed output tiles are flushed back to the host memory. A lightweight controller manages double-buffered DMA transfers and triggers interrupts to the CPU. MatrixFlow supports multiple data precisions (INT8/16/32, FP16/32) with configurable PE logic and buffer size. Across configurations, the architecture maintains the same principle: minimal local storage, efficient memory streaming, and alignment with system-level constraints, paving the way for the integrated design discussed in Section~\ref{subsec:gem5-accesys}.

\vspace{-0.25cm}
\subsection{System Integration and Gem5-AcceSys Framework}
\label{subsec:gem5-accesys}
To evaluate MatrixFlow in a realistic system context, we integrate the accelerator into a full-system gem5 simulation using our Gem5-AcceSys framework. The simulated host is an ARM-based multicore CPU with private caches and a shared last-level cache (LLC) connected to DDR3 main memory. The MatrixFlow accelerator attaches as a peripheral via a PCIe link on the system bus, enabling it to issue memory requests and DMA transfers like a standard I/O device. This loosely coupled approach at the interconnect level requires no modifications to the CPUs and leverages standard interfaces (PCIe) for seamless integration.
Gem5-AcceSys augments the simulator with key SoC components to support the accelerator. Notably, it includes a system memory management unit (SMMU) for the accelerator’s virtual-to-physical address translation and a multi-channel DMA engine for streaming data between host memory and the accelerator’s local buffers. These features allow the accelerator to share the host’s virtual address space and perform high-throughput data transfers, closely emulating a real deployment. We introduced these enhancements in response to MatrixFlow design requirements (e.g., virtual memory support and efficient data streaming), underscoring the iterative codesign of the hardware and the system model.
Gem5-AcceSys also captures realistic interconnect and software behavior. For example, the PCIe 3.0x4 interface of the accelerator is modeled with proper link latency, bandwidth limitations, and contention on the host memory bus. The framework runs a full Linux OS with a lightweight kernel driver to manage the accelerator, mirroring a real software stack. This driver sets up DMA buffers for I/O, issues commands to the accelerator (by writing to its control registers) and handles interrupts on completion, just as a real system driver would.

Our platform supports three modes for the accelerator’s memory access (Fig.~\ref{fig:designFrame}):

\textbf{DM (Direct-Memory)}—\emph{arrows (3,\,5)}: The accelerator’s DMA issues host virtual-address requests; the SMMU performs translation; traffic traverses PCIe/host interconnect and reaches DRAM via (5). DM maximizes streaming bandwidth for large transfers but bypasses the CPU LLC, so software must explicitly flush/invalidate to ensure coherence.

\textbf{DC (Direct-Cache)}—\emph{arrows (2,\,4,\,5)}: Accelerator load/stores enter the coherent LLC; hits return at cache latency, while misses traverse the on-chip MemBus (2) to the memory controller (4) and DRAM (5). DC is favorable when tiled working sets exhibit LLC locality (e.g., reuse across neighboring tiles or pipeline stages).

\textbf{DevMem (Device-side)}—\emph{arrow (6)}: The accelerator reads/writes its on-card SRAM/DRAM via a local controller, avoiding PCIe for resident tiles. This eliminates host-memory traffic during GEMM-heavy phases, but CPU-resident pre/post stages (e.g., packing, activation, or control) may incur PCIe crossings.

We use these exact arrow sequences in Sect.~\ref{sec:results} to interpret performance trends (e.g., Fig.~\ref{fig:device_vs_host_memory}: host-memory with fast PCIe vs.\ DevMem residence) and to quantify when DC’s locality beats DM’s raw bandwidth.

The runtime partitions matrices into 4\,KB tiles
(\textsc{Int8}~$16{\times}256$, \textsc{Fp16}/\textsc{Int16}~$16{\times}128$, \textsc{Fp32}/\textsc{Int32}~$16{\times}64$).
The driver pins/maps the pages via the SMMU, enqueues \texttt{read(A)}, \texttt{read(B)}, and \texttt{write(C)} DMA descriptors, and rings a doorbell (MMIO).
The DMA then fetches A/B along the datapath of the selected mode—
\textbf{DM} \emph{(3,\,5)}, \textbf{DC} \emph{(2,\,4,\,5)}, or \textbf{DevMem} \emph{(6)}—to fill the input tile buffers.
While the 16${\times}$16 systolic array (SA) consumes the current A/B tiles, the DMA prefetches the next A/B tiles and drains the previous C tile. Upon completion, the accelerator raises an MSI interrupt; the driver recycles descriptors and the runtime advances to the next tile.
This control-plane sequence aligns with the data pipeline in Fig.~\ref{fig:matrix_pipeline}, and—by referencing the Fig.~\ref{fig:designFrame} arrows—makes the hardware–system integration explicit (e.g., \emph{LLC hits on DC} versus \emph{raw streaming on DM}). We evaluate these three paths in Sect.~\ref{sec:results} to quantify how interconnect and memory placement shape end‑to‑end throughput.

In general, this integrated design ensures that data movement overlaps with computation. As one tile of a matrix is processed in the accelerator systolic array, the next tile is fetched (from memory or cache) and a previous output tile is written back, creating a pipeline that keeps communication and computation in parallel. In effect, what would traditionally be a sequence of discrete transfers becomes a continuous stream of data flowing from memory to the accelerator and back. This holistic co-design of MatrixFlow and Gem5-AcceSys means every hardware decision is mirrored in the system model and each system insight feeds into the hardware design, ensuring the accelerator and its system context are optimized together rather than in isolation.
\vspace{-0.25cm}
\subsection{Dataflow Optimization and Software Mapping}
\label{subsec:dataflow}
\begin{figure}[htbp]
\vspace{-0.25cm}
  \centering
  \includegraphics[width=1.0\linewidth]{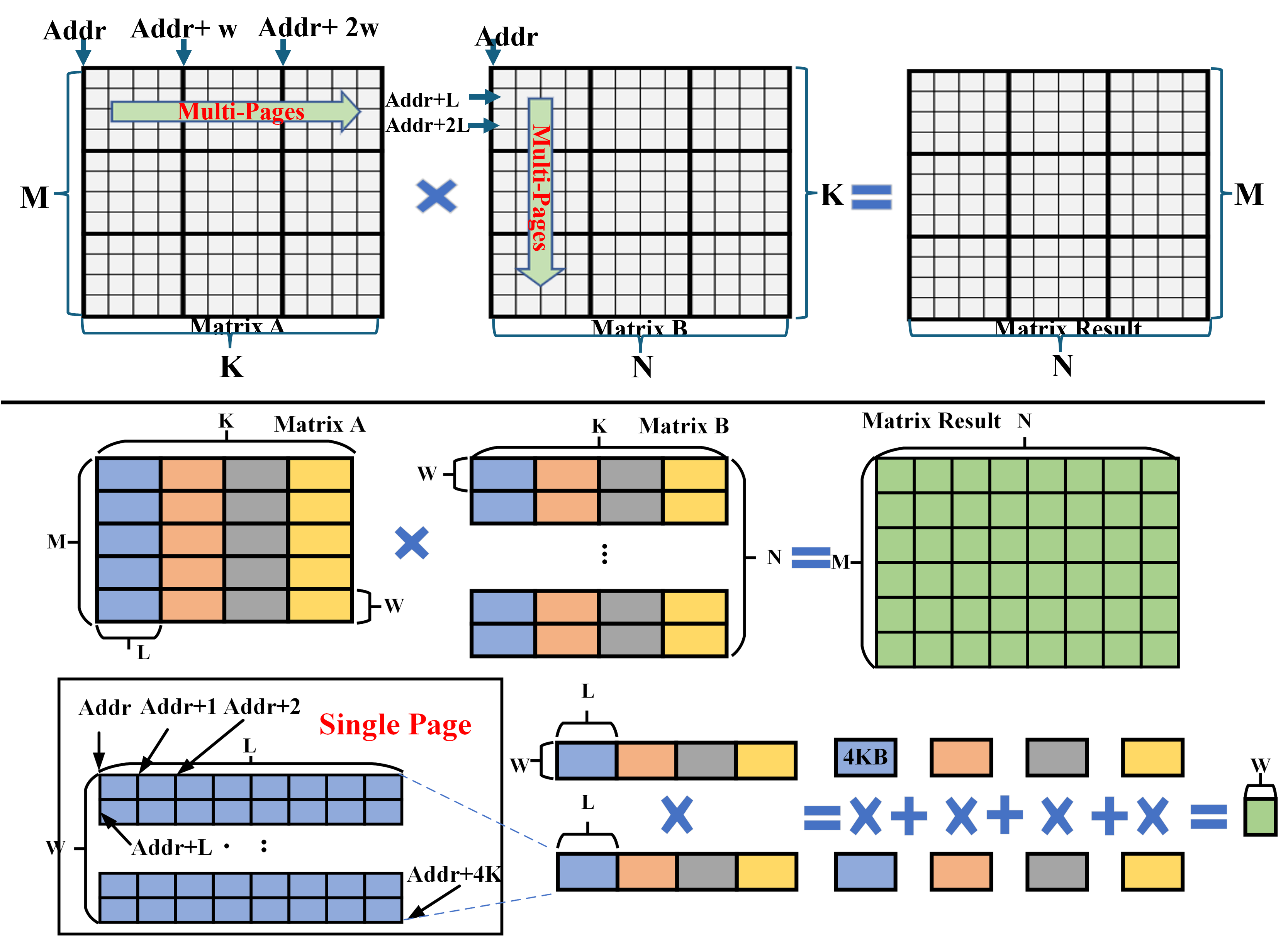}
  \caption{Conventional matrix blocking and multiplication (top) and our method for GEMM (bottom)~\cite{Liu2025}}
  \Description{Conventional Matrix blocking and multiplication and our method for GEMM}
  \label{fig:matrix_mul_combined}
  \vspace{-0.15cm}
\end{figure}

We choose 4 KB tiles to align each DMA burst with a single page translation in the SMMU and to match multiples of 16 for the 16×16 systolic array. This reduces PCIe packetization waste and TLB pressure while keeping the SA fully occupied with minimal on‑chip SRAM (two 4 KB inputs + one 4 KB output).
Naïve GEMM implementations in accelerators often traverse matrix A in row-major order and matrix B in column-major order. As shown in Figure~\ref{fig:matrix_mul_combined} (top), this mismatch causes B’s data to be accessed in non-contiguous, strided patterns: entire columns of B are scattered across many pages, leading to poor cache locality, frequent TLB misses, and inefficient use of PCIe bandwidth (many small transfers). In short, a straightforward mapping of the matrix multiplication would be bottlenecked by memory access overheads rather than by the accelerator’s compute capabilities.

We address this issue by co-optimizing the data layout and the accelerator dataflow. The runtime partitions each large matrix into page-aligned 4KB tiles (e.g., 64×64 INT8 elements per tile). Each tile of matrix A is stored in a row-major format, and each tile of matrix B is stored contiguously in a row-striped manner (i.e., by rows rather than true columns) to avoid strided accesses. Figure~\ref{fig:matrix_mul_combined} (bottom) illustrates this block-based tiling scheme. With this organization, every tile can be fetched with a single DMA burst spanning one full page, eliminating fragmented transfers and reducing address-translation overhead.

In the software runtime, the GEMM is executed tile by tile using this blocked layout (Algorithm~\ref{alg:block_matrix_multiplication}). The nested loops iterate over tiles A and B, accumulating partial products into each output tile and writing back each completed W×W result block before proceeding to the next. This tile-by-tile processing effectively transforms the matrix multiplication into a series of streaming page-sized data transfers and computations, instead of numerous fine-grained element-wise accesses.

\begin{algorithm}
\caption{Optimized Block Matrix Multiplication}
\label{alg:block_matrix_multiplication}
\begin{algorithmic}[1]
    \STATE \textbf{function} \textsc{BlockMatrixMultiply}($A, B, M, N, K$)
    \STATE $Res \gets$ InitializeMatrix($M, N$)
    \STATE Divide $A$ and $B$ into blocks of size $W \times L$
    \FOR{$i = 0$ to $M / W - 1$}
        \FOR{$j = 0$ to $N / W - 1$}
            \STATE $Res_{block} \gets$ ZeroMatrix($W, W$)
            \FOR{$k = 0$ to $K / L - 1$}
                \STATE $A_{block} \gets$ GetBlock($A, i, k$)
                \STATE $B_{block} \gets$ GetBlock($B, j, k$)
                \STATE $Res_{block} \gets$ MultiAcc($A_{block}, B_{block}, Res_{block}$)
            \ENDFOR
            \STATE SetBlock($Res, i, j, Res_{block}$)
        \ENDFOR
    \ENDFOR
    \STATE \textbf{return} $Res$
\end{algorithmic}
\end{algorithm}

\begin{figure}[htbp]
  \centering
  \includegraphics[width=1.0\linewidth]{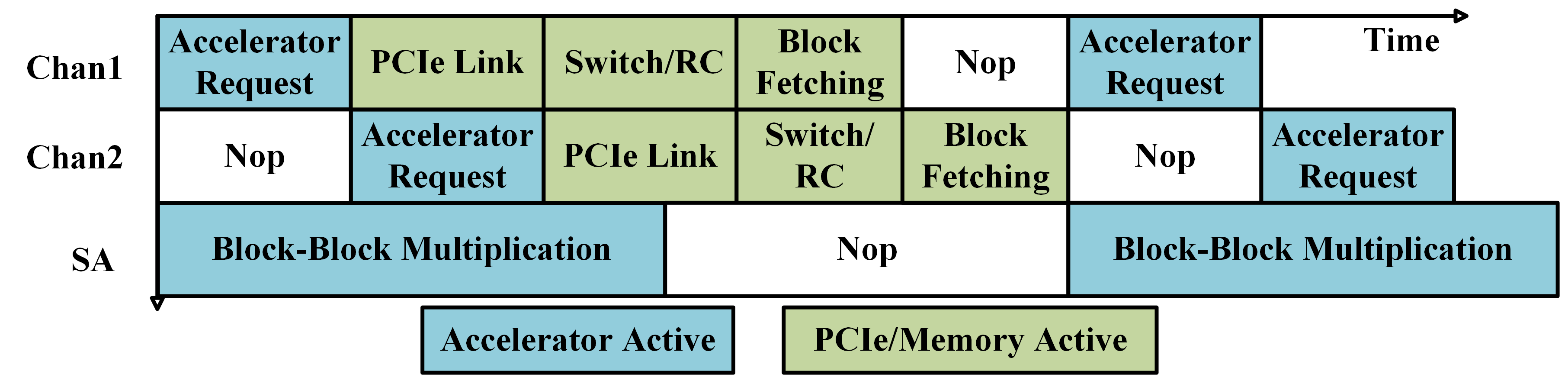}
  \caption{Matrix multiplication pipeline~\cite{Liu2025}}
  \Description{Matrix multiplication pipeline}
  \label{fig:matrix_pipeline}
  \vspace{-0.25cm}
\end{figure}

The benefits of this co-designed dataflow are multi-fold. First, it maximizes contiguous data transfers over the PCIe and memory system: by always requesting full-page chunks, we achieve high payload utilization of each PCIe transaction and reduce per-transfer overhead. Second, it greatly reduces address translation overhead – each 4KB block incurs at most one page translation (TLB lookup), as opposed to many if the same amount of data were scattered across pages. Third, blocking improves cache utilization and spatial locality. Even in DC mode (cache-coherent access), the 4KB tiling means that once a block of B is brought into the LLC, the accelerator can reuse all bytes in that block for many 16×16 computations before moving to the next, instead of thrashing on uncached column accesses. Fourth, by matching the tile dimensions to the systolic array’s 16×16 PE dimensions, we avoid internal fragmentation in the processing elements – the full 16×16 array can be kept busy on one block at a time with minimal padding or wasted computation. And finally, this block-based scheme enables a pipelined execution: while one tile is being processed by the systolic array, the next tile is prefetched into input buffers and the previous result tile is written back, creating an overlapping producer–consumer pipeline across tiles. Figure~\ref{fig:matrix_pipeline} illustrates this matrix multiplication pipeline, where reading A, reading B, calculating the systolic array, and writing back C are all overlapped for successive blocks. As a result, the accelerator is rarely idle – computation and data transfer proceed concurrently for different tiles, which sustains a high overall throughput. This dataflow-driven software mapping is a pivotal factor in MatrixFlow’s performance gains: By co-optimizing how data is laid out and fetched with the hardware capabilities, we free the host CPU from managing fine-grained data movement and allow the accelerator to run at full speed with minimal memory stall time.

In summary, the system-accelerator co-design extends beyond hardware integration to the algorithm level: the 4KB blocked matrix multiplication method ensures that MatrixFlow operates on the transformer’s large matrices in a cache-friendly, stream-friendly manner, significantly increasing end-to-end inference throughput.
Each of these innovations – from block segmentation and page-aligned tiling to horizontal reordering of data – contributes to an overall design that bridges the gap between computational efficiency and system-level data movement, a balance that is essential for high-performance transformer inference.

\section{Experimental Setup and Methodology}
\label{sec:method}
All experiments are carried out on \textbf{Gem5-AcceSys} with full-system simulation based on Linux.
\vspace{-0.25cm}
\subsection{Host–Accelerator Platform}
\label{subsec:acceleplatform}
Table~\ref{tab:platform} summarizes the baseline system.  Unless otherwise stated, the accelerator streams data from host DDR3 over a Gen6~$\times$16 link (128\,GB\,s$^{-1}$ bidirectional peak).  In Section~\ref{sec:results} we sweep (i)~link width/speed and (ii)~memory technology (DDR4/5, GDDR6, on-card HBM2).

\begin{table}[htbp]
\vspace{-0.15cm}
  \centering\small
  \caption{Baseline host–accelerator configuration.}
  \vspace{-0.1cm}
  \label{tab:platform}
  \begin{tabular}{lcc}
    \toprule
    \textbf{Component} & \textbf{Parameter} & \textbf{Value}\\
    \midrule
    CPU\;(\textit{host})      & core\;@\;freq & ARMv8 OoO @ 1 GHz\\
    L1 I/D cache             & size & 32 / 64 kB, 4-way\\
    LLC                      & size & 2 MB, 16-way, inclusive\\
    Main memory (host)       & tech / size & DDR3-1600, 4 GB\\
    PCIe interconnect        & spec & Gen6 $\times$16 (64 GT s$^{-1}$/lane)\\
    IOMMU (SMMU)             & page size & 4 kB (TLB 64 entries)\\
    Accelerator (MatrixFlow) & array & 16×16 MACs, 1 GHz\\
                             & on-chip SRAM & 3×4 kB tiles\\
    DMA channels             & read / write & 2 / 2, 1024-B bursts\\
    \bottomrule
  \end{tabular}
  \vspace{-0.1cm}
\end{table}

\subsection{Benchmarks and Workloads}
\label{subsec:benchmarks}
To evaluate the performance of our system–accelerator co‑design, we use two main types of workloads:
\textbf{Micro-kernels:}  Square GEMMs from $64\times64$ to $2048\times2048$ (INT8/16 and FP16/32) under different configurations (e.g., compute- vs.\ I/O-bound).
\textbf{Transformer inference:}  For transformer-based workloads, we run:
\begin{itemize}[leftmargin=*]
  \item \emph{BERT-Base} (12 layers, 110 M params, seq.\ length 128)  
  \item \emph{BERT-Large} (24 layers, 340 M params, seq.\ length 128)
  \item \emph{ViT-Base/16} (12 layers, 86 M params, $224^{2}$ image)
  \item \emph{ViT-Large/16} (24 layers, 307 M params, $224^{2}$ image)
\end{itemize}

These are full transformer encoder models rather than synthetic kernels. BERT‑Base/Large are widely‑used backbones in deployed NLP services such as text classification and question answering. ViT‑Base/Large serve as standard vision transformer backbones not only for ImageNet classification but also for dense prediction tasks (e.g., \cite{ranftl2021_vit_dense_prediction,chen2023_vit_adapter}) and for industrial visual inspection and defect‑detection systems. Because these workloads share the same stack of multi‑head self‑attention and feed‑forward blocks as many other transformer models, the resulting GEMM/non‑GEMM mix and system‑level behavior are representative of a broad class of transformer inference workloads.

All GEMM calls inside attention and feed-forward blocks are off-loaded to MatrixFlow; element-wise and soft-max stages are executed on the host CPU.

\subsection{Hardware Implementation, Power and Energy Modeling}
\label{subsec:impl-power}
We synthesized a family of MatrixFlow systolic arrays (SAs) to obtain post-synthesis PPA: \{4$\times$4, 16$\times$16\} arrays across \{INT8/16/32, FP8/16/32\}. Fixed-point designs target 1.0\,GHz; floating-point designs meet timing at 1.66\,ns ($\approx$0.60\,GHz). Table~\ref{tab:synth-ppa} reports total cell area and dynamic power for the Systolic-Array (compute only).
\begin{table}[htbp]
\centering
\caption{Post-synthesis systolic-array PPA and derived compute-only efficiency (MAC = 2 ops).}
\vspace{-0.1cm}
\label{tab:synth-ppa}
\resizebox{\columnwidth}{!}{
\begin{tabular}{lrrrrr}
\toprule
\textbf{Variant} & \textbf{Freq} & \textbf{Area ($\mu$m$^2$)} & \textbf{Power (mW)} & \textbf{Peak (GOPS)} & \textbf{GOPS/W (TOPS/W)} \\
\midrule
INT8 4$\times$4     & 1.0\,GHz  & 16{,}186     & 7.464   & 32.0    & 4{,}287\;(\,4.29\,) \\
INT8 16$\times$16   & 1.0\,GHz  & 186{,}875    & 84.550  & 512.0   & 6{,}056\;(\,6.06\,) \\
\midrule
INT16 4$\times$4    & 1.0\,GHz  & 24{,}989     & 11.813  & 32.0    & 2{,}709\;(\,2.71\,) \\
INT16 16$\times$16  & 1.0\,GHz  & 397{,}558    & 149.419 & 512.0   & 3{,}427\;(\,3.43\,) \\
\midrule
INT32 4$\times$4    & 1.0\,GHz  & 73{,}483     & 33.302  & 32.0    &    961\;(\,0.96\,) \\
INT32 16$\times$16  & 1.0\,GHz  & 1{,}163{,}841 & 392.978 & 512.0   & 1{,}303\;(\,1.30\,) \\
\midrule
FP8 4$\times$4     & 0.60\,GHz & 8{,}806      & 2.251   & 19.2    & 8{,}530\;(\,8.53\,) \\
FP8 16$\times$16   & 0.60\,GHz & 142{,}816    & 34.557  & 307.2   & 8{,}890\;(\,8.89\,) \\
\midrule
FP16 4$\times$4    & 0.60\,GHz & 22{,}802     & 5.580   & 19.2    & 3{,}441\;(\,3.44\,) \\
FP16 16$\times$16  & 0.60\,GHz & 363{,}805    & 83.655  & 307.2   & 3{,}672\;(\,3.67\,) \\
\midrule
FP32 4$\times$4    & 0.60\,GHz & 62{,}693     & 16.938  & 19.2    & 1{,}134\;(\,1.13\,) \\
FP32 16$\times$16  & 0.60\,GHz & 1{,}032{,}820 & 258.173 & 307.2   & 1{,}190\;(\,1.19\,) \\
\bottomrule
\end{tabular}}
\end{table}

MatrixFlow uses only three \(4\,\text{kB}\) on-chip SRAM blocks (A/B/C tiles, \(12\,\text{kB}\) total logical capacity), whereas many systolic-array designs provision hundreds of kB of local scratchpad. This small-SRAM choice is deliberate: most model parameters reside in host or device DRAM, and the accelerator relies on streaming to keep the array busy.
Table~\ref{tab:synth-ppa} reports the synthesized area and power for our \(4{\times}4\) and \(16{\times}16\) SAs across INT/FP precisions, showing that INT8 offers the best energy efficiency. Compared with a Gemmini-like tightly coupled SA and the e-GPU tensor core, MatrixFlow achieves similar or better TOPS/W and compute density while using an order-of-magnitude less on-chip SRAM.

Compared to Gemmini and the e-GPU: Gemmini is a tightly coupled RoCC co-processor that shares the CPU memory hierarchy and uses software-managed local memories (scratchpad + accumulator) to exploit on-chip reuse; typical designs provision hundreds of kB on-chip.\,By contrast, MatrixFlow is loosely coupled and deliberately minimal-SRAM: only three page-sized SRAMs (A/B/C = 3$\times$4\,KB) are used as staging FIFOs, with high-throughput DMA streaming aligned to 4\,KB OS pages (§3.1–§3.3). This co-design shifts storage/coordination into the system (PCIe + host DRAM + SMMU) rather than silicon area. In edge GPUs, high absolute throughput is achieved via large on-die caches/register files and wide device memory, but sustaining efficiency often depends on large on-chip state and kernel residency. MatrixFlow takes the opposite path: a small-footprint SA fed by page-aligned DMA bursts, trading large scratchpads for sustained streaming to keep utilization high.

Our post-synthesis results show that the \(16{\times}16\) INT8 SA at \(1\,\text{GHz}\) reaches \(512\,\text{GOPS}\) and about \(6.1\,\text{TOPS/W}\) for the compute array alone; including the \(12\,\text{kB}\) SRAM dynamics still yields roughly \(3.6\,\text{TOPS/W}\), reinforcing that the system is primarily bounded by memory bandwidth and PCIe transaction efficiency rather than raw SA peak.

\section{Results and Design Space Exploration for Transformer Accelerator}
\label{sec:results}
\subsection{Performance Characterization of MatrixFlow Accelerator}
\label{subsec:matrixflow-character}
\begin{figure}[htbp]
  \centering

  \begin{subfigure}[b]{0.48\linewidth}
    \centering
    \includegraphics[width=\linewidth]{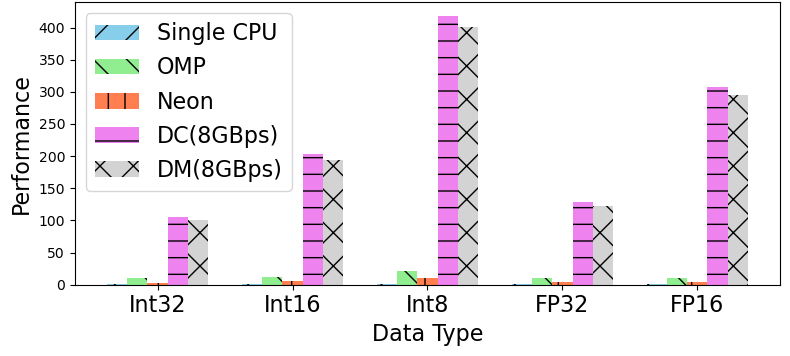}
    \caption{Performance Comparison on Precision~\cite{Liu2025}}
    \label{fig:perf_sgl}
  \end{subfigure}
  \hfill
  \begin{subfigure}[b]{0.48\linewidth}
    \centering
    \includegraphics[width=\linewidth]{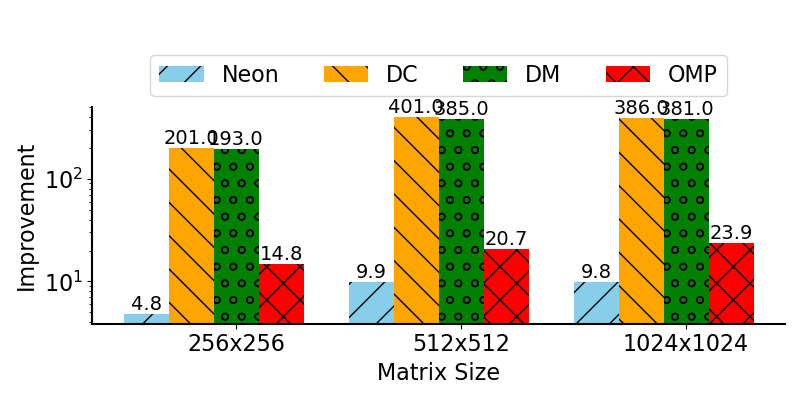}
    \caption{Performance Comparison on Matrix Size~\cite{Liu2025}}
    \label{fig:perf_mixed}
  \end{subfigure}
  \vspace{-0.15cm}
  \caption{Matrix-multiplication performance comparison on (a) different precision; (b) different matrix size.}
  \Description{Matrix-multiplication performance comparison}
  \label{fig:matrix_perf}
  \vspace{-0.25cm}
\end{figure}

\begin{figure}[htbp]
  \vspace{-0.25cm}
  \centering
  \includegraphics[width=0.5\linewidth]{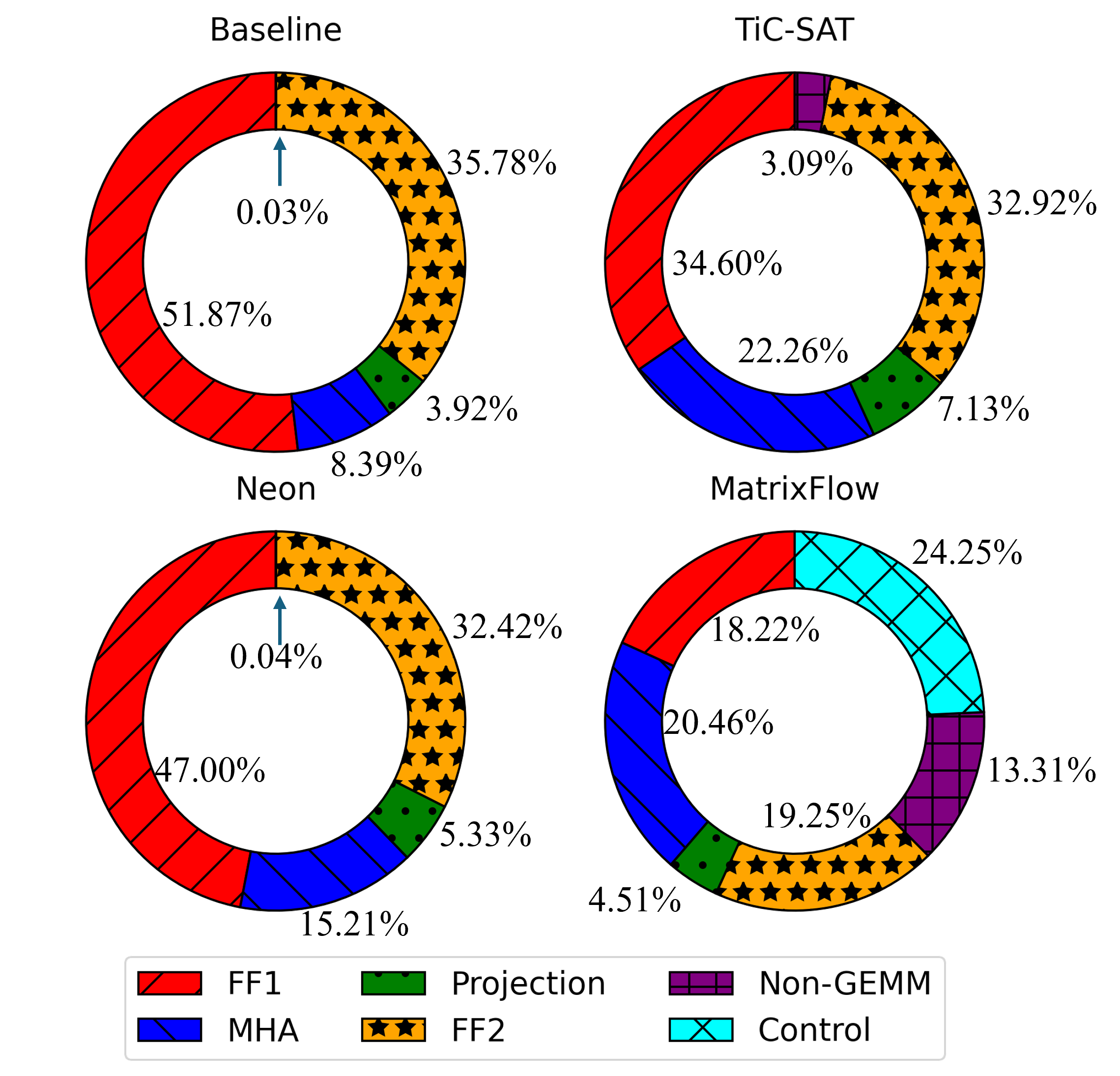}
  \vspace{-0.15cm}
  \caption{Runtime Analysis~\cite{Liu2025}}
  \Description{Runtime Analysis}
  \label{fig:runtimeAna}
  \vspace{-0.25cm}
\end{figure}

This section evaluates the performance of MatrixFlow both at the computation kernel level (gemm alone throughput) and at the full transformer inference level, to demonstrate the advantages of the accelerator. We compare against a single-core CPU baseline and prior accelerator designs. Figures~\ref{fig:matrix_perf} (a) and (b) present the raw GEMM throughput of MatrixFlow across different data precisions and matrix sizes, while Figure~\ref{fig:runtimeAna} provides a breakdown of the end-to-end inference run time by operation, contrasting MatrixFlow with the CPU and other accelerators.

Figure~\ref{fig:matrix_perf}(a) highlights the raw GEMM throughput across data precisions and matrix sizes, as well as a breakdown of runtime for complete transformer models, contrasting MatrixFlow against CPU and other accelerator baselines. It studies matrix multiplication performance for various data types (512×512 matrix) on different platforms: single-thread CPU (baseline), 256-thread CPU (OMP), ARM Neon (SIMD), and MatrixFlow accelerator (with direct-cache DC and direct-memory DM). 
Higher bars indicate greater throughput or speed-up. The MatrixFlow accelerator achieves dramatically higher GEMM throughput than the CPU on all precisions. 
MatrixFlow sees the greatest speedup with FP16 data, because the baseline CPU lacks native FP16 support and must resort to slow software emulation. 
In contrast, the CPU-based implementations (even with Neon SIMD or 256 threads) gain performance as precision decreases (int8 > int32), but their absolute speedups remain much lower. The 256-thread CPU reaches only on the order of 20–30× speed-up on this 512×512 GEMM, far below MatrixFlow’s hundreds-fold improvement.
Across all data types, MatrixFlow sustains the highest throughput – demonstrating the advantage of a specialized systolic array accelerator with an optimized dataflow, compared to general-purpose processors.

Fig.~\ref{fig:matrix_perf}(b) describes the GEMM speed-up with different matrix size using Int8 as desired data type for MatrixFlow compared to CPU baselines. Bars show improvement factors (log scale) relative to single-core execution, for three matrix sizes. 
direct-cache (DC) and direct-memory (DM) denote MatrixFlow’s access modes; OMP is a 256-thread CPU; Neon is single-core SIMD. 
MatrixFlow exhibits extremely large speedups that grow with problem size, reaching ~400× at a 512×512 matrix (in direct-cache mode). In these int8 tests, MatrixFlow's direct cache (DC) mode slightly outperforms the direct memory (DM) mode for the largest matrices (e.g., 1024×1024), due to better use of the CPU’s cache hierarchy in the DC setup; however, both modes vastly outperform the CPU baselines.
The 256-thread CPU implementation saturates at only ~20–25× speed-up, and even an optimized single-core Neon SIMD achieves <10×, at these matrix sizes. 
This stark contrast highlights MatrixFlow’s superior scaling efficiency: As matrix dimensions grow, MatrixFlow continues to exploit data reuse and high memory bandwidth, while the CPU quickly hits its limits in memory bandwidth and parallelism. 

Fig.~\ref{fig:runtimeAna} illustrates the breakdown of inference runtime by major operation on four platforms: (a) a single-core ARM CPU baseline, (b) a CPU with Neon SIMD optimizations, (c) the TiC-SAT~\cite{Amirshahi2023} tightly-coupled accelerator, and (d) the MatrixFlow accelerator. The baseline is a single ARM core, no Neon SIMD; ‘Neon-optimized CPU’ uses ARM Neon intrinsics for GEMM to show the benefit of SIMD.
Each donut chart shows the percentage of total runtime spent in key operations: Feed-Forward layers 1 and 2 (FF1, FF2), Multi-Head Attention (MHA), output projection, non-GEMM operations, and control/communication overhead. Each segment indicates the percentage of time spent in a specific component of the transformer workload (e.g., GEMM computations, feed-forward layers, softmax/normalization, or host-accelerator communication overheads).

We measure per-class times in gem5/Gem5-AcceSys by tagging each kernel launch and DMA episode with a class label, and then aggregating wall-clock time by class: FF1/FF2 (MLP GEMMs), MHA (attention GEMMs), Projection (QKV/out linears), Non-GEMM (softmax, layer norm, activations), and Control (CPU orchestration, descriptor setup, SMMU/IOMMU page walks). Unless stated otherwise, numbers are the mean over runs of ViT-Base (batch=1, 224×224).
The baseline CPU is almost entirely dominated by matrix computations. GEMM-based layers consume ~99\% of the inference time (with the two feed-forward layers alone accounting for $\approx$87.7\%). Offloading computation to Neon (SIMD) reduces this somewhat, but GEMM operations still remain the bottleneck. With accelerators, the profile shifts significantly. In MatrixFlow, the heavy GEMM layers are drastically accelerated, so other costs become more visible: about 13\% of time is spent in non-GEMM parts of the workload (e.g. softmax, normalization), and importantly a control overhead of ~24\% emerges from the loosely-coupled design (PCIe/DMA transfers and driver interaction).
By shifting the primary bottleneck from CPU compute to PCIe/DMA communication, MatrixFlow deliberately accepts an additional \(\approx24\%\) communication overhead, yet still outperforms TiC-SAT overall.
In MatrixFlow, the primary bottleneck has been shifted from the compute-intensive GEMM layers to the I/O subsystem. TiC-SAT, being tightly integrated, incurs negligible PCIe/DMA overhead, but it still spends a large fraction of runtime in the feed-forward (FF) layers and other non-GEMM operations, even though those are accelerated, because their performance is limited by the CPU’s internal bandwidth and data reuse constraints.
Notably, by offloading the large matrix multiplications more efficiently, MatrixFlow actually reduces the relative time spent in the feed-forward layers compared to TiC-SAT, even though MatrixFlow incurs communication overhead. 
\vspace{-0.25cm}
\subsection{Microbenchmark Analysis}
\label{subsec:microbenchana}
\begin{figure}[htbp]
\vspace{-0.25cm}
  \centering
  \includegraphics[width=1.0\linewidth]{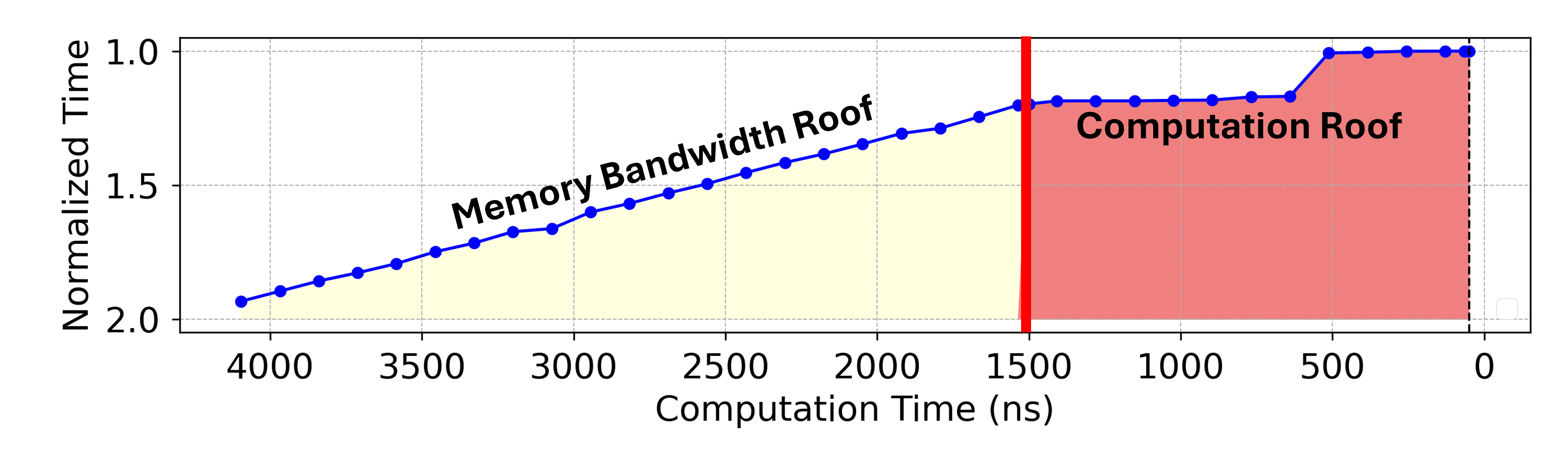}
  \vspace{-0.75cm}
  \caption{Roofline Model of the Accelerator System~\cite{Liu2025AcceSys}}
  \Description{Roofline Model of the Accelerator System}
  \label{fig:roofline}
  \vspace{-0.25cm}
\end{figure}

We begin by assessing the platform’s theoretical performance limits with a roofline-style model. Figure~\ref{fig:roofline} illustrates the system’s execution time as a function of ideal accelerator compute throughput. Initially, increasing compute capability yields nearly linear speedups, but beyond a certain point the curve flattens as the memory bandwidth “roof” is reached and additional compute can no longer improve performance. This defines an upper bound on useful accelerator speedup and underscores the need to balance computation and data movement in a co-designed system. 
To identify the practical bottlenecks in our accelerator, we conducted microbenchmark experiments focusing on three key factors: (i) the PCIe interface bandwidth and transaction packet size, (ii) the memory technology and data placement (device-side vs. host-side memory), and (iii) the address translation overhead. We used a representative 2048×2048 matrix multiplication to isolate each factor’s impact, as detailed below.

\vspace{-0.25cm}

\subsubsection{PCIe Bandwidth and Packet Size}

\begin{figure}[htbp]
\vspace{-0.25cm}
  \centering
  \includegraphics[width=0.7\linewidth]{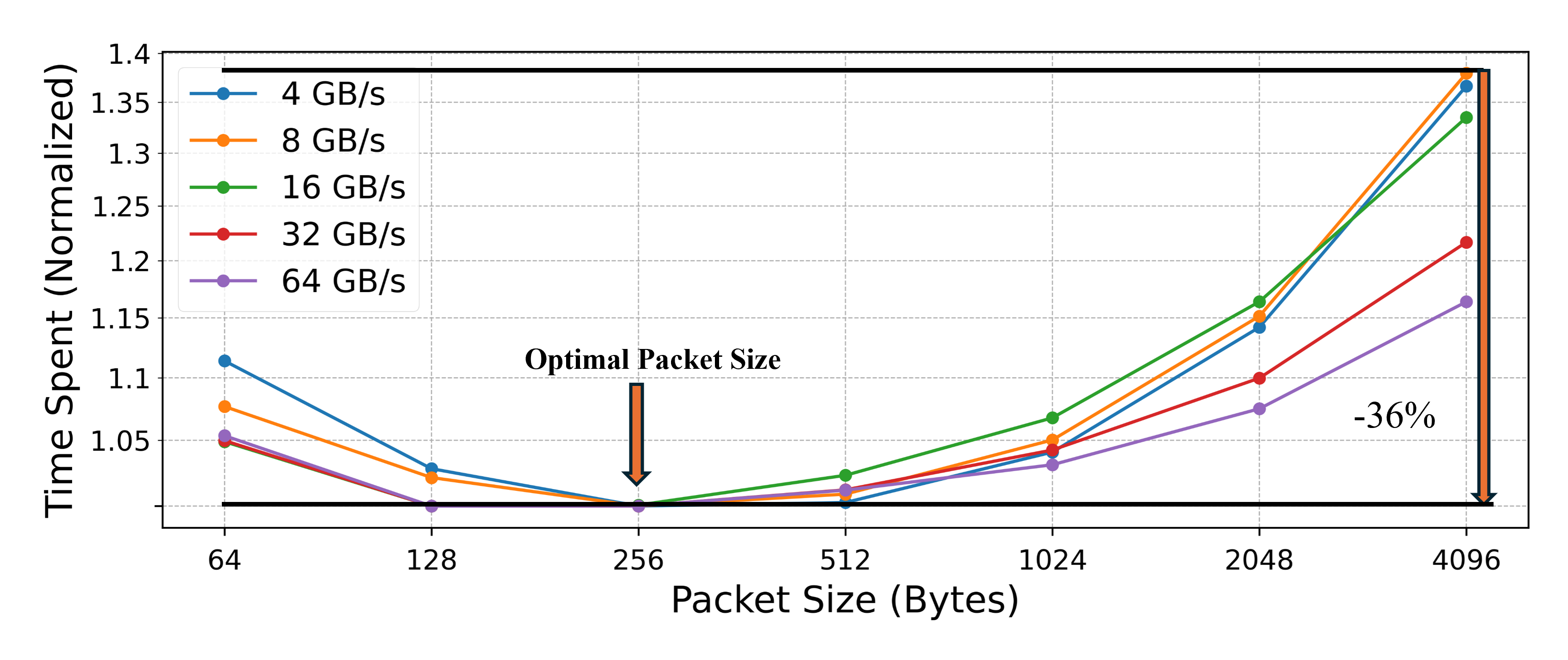}
  \vspace{-0.25cm}
  \caption{Execution Time under different packet sizes for different PCIe bandwidth~\cite{Liu2025AcceSys}.}
  \Description{Execution Time under different packet sizes for different PCIe bandwidth}
  \label{fig:GEMM_packet}
  \vspace{-0.25cm}
\end{figure}

As expected, increasing PCIe transfer throughput dramatically reduces execution time. In our experiments, scaling the PCIe link from a minimal configuration (2 lanes at 2 Gbps) up to a high-end link (16 lanes at 16 Gbps) yielded an ~11× speedup for the 2048×2048 GEMM workload. However, beyond raw bandwidth, the size of the PCIe transaction packet also affects efficiency (Figure~\ref{fig:GEMM_packet}). Very small transfers (64B) incurred about 12\% longer execution time than the optimal size due to higher per-packet overhead, while extremely large packets (4096B) caused pipeline stalls in the PCIe subsystem, leading to up to 36\% slowdown at lower link speeds. We observed an optimal transaction size of around 256B across all tested bandwidths, which balances transfer overhead and pipeline utilization. Thus, both a high-bandwidth PCIe interface (e.g., ×16 Gen4/Gen5) and a properly tuned packet size ($\approx$256B) are critical for minimizing data transfer bottlenecks.

\subsubsection{Memory Technology and Location}

Next, we analyze the impact of the memory subsystem. Our platform supports both device-side memory (dedicated DRAM attached directly to the accelerator) and host-side memory (system memory accessed over PCIe), with configurable models for common DRAM types. We evaluated five memory technologies, DDR3, DDR4, DDR5, GDDR6, and HBM2, each in two scenarios: attached directly to the accelerator vs. accessed via the host interface. Table~\ref{tab:mem_config} summarizes the key parameters of these memory configurations (e.g., per-channel data width and bandwidth ranging from 12.8GB/s for DDR3 up to 64GB/s for HBM2).

\begin{table}[htbp]
\vspace{-0.15cm}
\caption{Memory Configuration}
\vspace{-0.1cm}
\centering
\begin{tabular}{lcccc}
\hline
\textbf{Component} & \textbf{Channel} & \textbf{Data width} & \textbf{Bandwidth} & \textbf{Data Rate} \\
\hline
DDR3 &  1 &  64 & 12.8 GB/s & 1600 MT/s \\
DDR4 &  1 &  64 & 19.2 GB/s & 2400 MT/s \\
DDR5 &  2 &  32 & 25.6 GB/s & 3200 MT/s \\
HBM2 & 2 &  128 & 64 GB/s & 2000 MT/s \\
GDDR6 & 2 &  64 & 32 GB/s & 2000 MT/s \\
\hline
\end{tabular}
\label{tab:mem_config}
\vspace{-0.1cm}
\end{table}

\begin{figure}[htbp]
  \vspace{-0.25cm}
  \centering
  \includegraphics[width=0.7\linewidth]{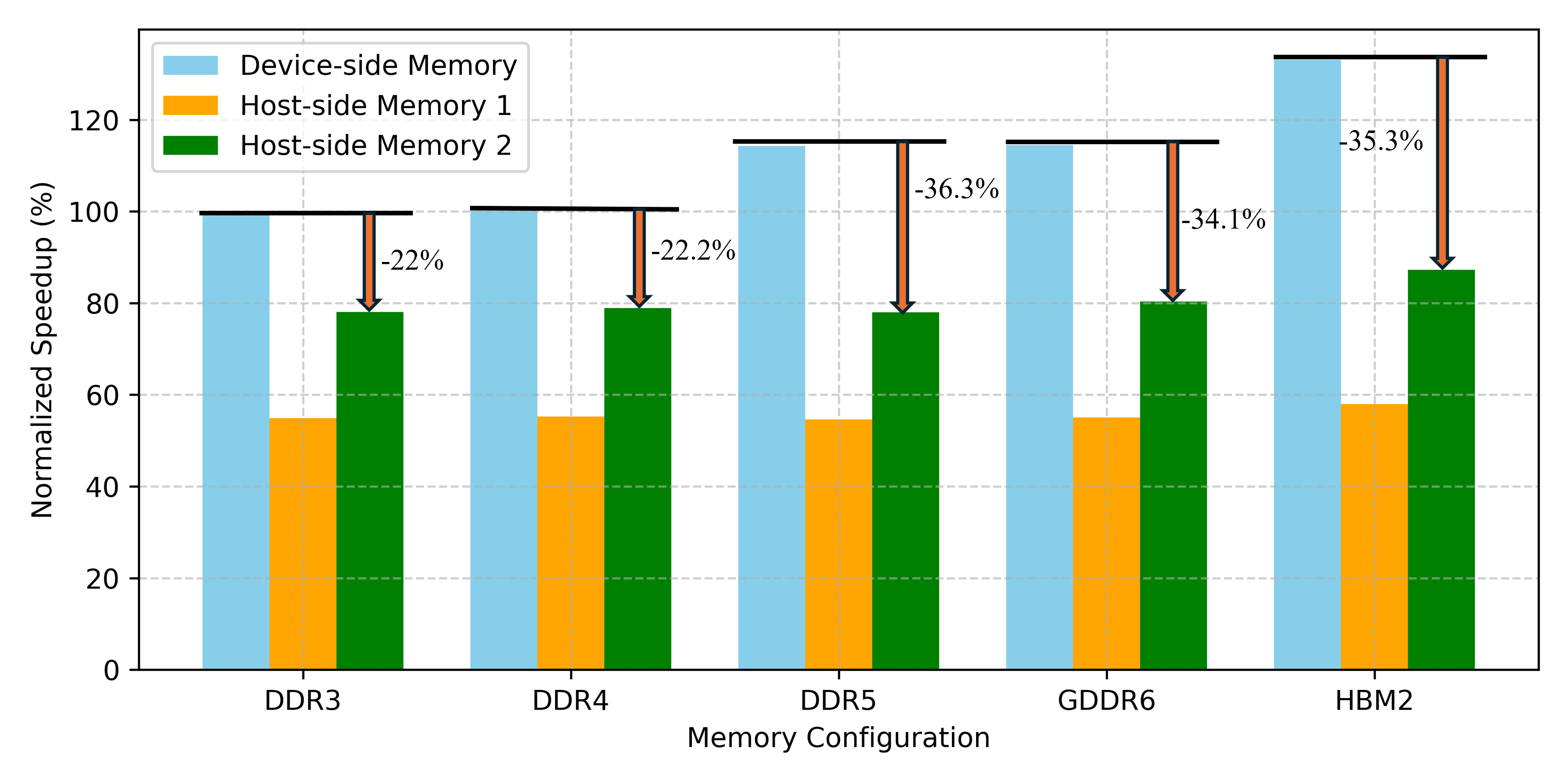}
  \vspace{-0.25cm}
  \caption{Impact of DRAM type and location~\cite{Liu2025AcceSys}}
  \Description{Impact of DRAM type and location}
  \label{fig:device_vs_host_memory}
  \vspace{-0.25cm}
\end{figure}

Figure~\ref{fig:device_vs_host_memory} compares GEMM performance for different memory types and placements. In all cases, using device-attached memory yields better performance than using host-accessed memory, and this gap grows wider for higher-bandwidth memories. For example, with HBM2 ($\approx$64GB/s per stack in our setup), the 2048×2048 matrix multiplication completes much faster when that HBM2 is directly on the accelerator, whereas accessing the same high-speed memory over the PCIe link is bottlenecked by the interconnect. Even a high-bandwidth PCIe (Gen5 ×16) cannot fully bridge the difference – it narrows the performance gap (by improving host-memory throughput) but still falls short of the device-attached case. These results highlight the importance of memory placement: to fully exploit fast memories like HBM2 or GDDR6 on accelerators, they must be integrated on the device side; otherwise, the benefits are limited by off-chip interface constraints.

In addition to memory placement, we performed sensitivity studies to separate the effects of memory bandwidth vs. latency on the GEMM workload. Keeping DRAM latency constant, we varied the available memory bandwidth over a wide range (up to 256GB/s). We found that performance improves rapidly with increased bandwidth up to around 50GB/s, beyond which returns diminish significantly – for example, increasing the memory bandwidth from 50GB/s to 256GB/s yielded only an extra ~1.7\% reduction in execution time. In contrast, maintaining fixed bandwidth and varying DRAM latency (from a baseline of 12 ns to 4 ns and up to 36ns) had only a minor effect: even tripling the latency increased execution time by merely ~4.9\% in total. This indicates that our transformer inference workload is far more bandwidth-bound than latency-bound in the memory subsystem. The implication for co-design is that investing in higher memory bandwidth (or better utilization of memory through improved dataflows and caching) will be more beneficial than aggressively minimizing latency, as long as latency remains within a reasonable range.

\subsubsection{Address Translation Overhead}

Finally, we examine the cost of virtual address translation in the full-system context. Modern accelerators often rely on an I/O Memory Management Unit (IOMMU) or the host CPU to handle address translation for device memory, which adds latency on TLB misses. Using Gem5-AcceSys, we quantified the translation overhead for progressively larger matrix multiplication workloads (increasing memory footprints); Table~\ref{tab:addrtrans} summarizes the results. For small matrices, the overhead is negligible (on the order of $\sim$1\% of runtime), but it rises sharply as the working set grows and TLB misses become more frequent. At the largest tested size (2048×2048, requiring $\approx$12,288 pages of memory), address translation accounts for about 6.5\% of the total execution time, with average multi-level page-walk latency exceeding 368 cycles. The mean per-access translation latency was lowest at an intermediate problem size ($\sim$1024×1024) but then jumped by roughly 5× at 2048×2048 once the TLB reach was exceeded. These results show that while address translation overhead is negligible for small models (or for accelerators with very large TLBs), it can become a non-trivial bottleneck for large models. In a co-designed system, it may therefore be necessary to support larger memory pages or incorporate more efficient TLB architectures when scaling to extremely large transformer models, in order to keep translation overhead manageable.

\begin{table}[htbp]
\vspace{-0.25cm}
\centering
\caption{Results with Larger Matrix Sizes~\cite{Liu2025AcceSys}}
\vspace{-0.1cm}
\label{tab:addrtrans}
\resizebox{\linewidth}{!}{%
\begin{tabular}{lcccccc}
\toprule
\textbf{Metric} & \textbf{64} & \textbf{128} & \textbf{256} & \textbf{512} & \textbf{1024} & \textbf{2048} \\
\midrule
Memory Footprint (Pages) & 12.0 & 48.0 & 192.0 & 768.0 & 3072.0 & 12288.0 \\
Translation Times & 3130 & 18470 & 142738 & 1082780 & 8593259 & 68430699 \\
Trans Mean Time & 23.42683 & 20.37948 & 13.87159 & 9.91735 & 10.478634 & 54.38005 \\
PTW Times & 15 & 54 & 227 & 1034 & 7675 & 479244 \\
PTW Mean Time & 176.6666 & 281.90740 & 265.255507 & 252.465184 & 294.609381 & 368.141137 \\
uTLB Lookup times & 2350 & 17690 & 137290 & 1081610 & 8586250 & 68423690 \\
uTLB Misses times & 195 & 862 & 8644 & 65808 & 731513 & 10416279 \\
Trans Overhead & 6.02\% & 1.87\% & 1.59\% & 1.30\% & 1.00\% & 6.49\% \\
\bottomrule
\end{tabular}}
\vspace{-0.25cm}
\end{table}

\vspace{-0.25cm}
\subsection{End-to-End Transformer Performance}
\label{sec:e2e-transformer}

To evaluate our accelerator in a complete application context, we measured end-to-end inference throughput on representative transformer models (BERT and ViT of various sizes) using the cache-coherent (DC) configuration (selected for its best GEMM performance in prior tests). The accelerator offloads all matrix multiplication operations, while the host CPU executes the remaining layers (e.g., Softmax, normalization, transpose). We compare against a single-core CPU baseline (as in Section~\ref{subsec:microbenchana}), an optimized CPU implementation using ARM NEON (with a custom data layout), and a multi-threaded CPU run with OpenMP on all cores. We also include two state-of-the-art accelerator frameworks for context: SMAUG (a gem5-based loosely-coupled DNN accelerator simulator) and TiC-SAT (a tightly-coupled systolic array accelerator with custom ISA extensions). These cover CPU-only, loosely coupled, and tightly coupled designs to put our co-designed MatrixFlow system in context.

\subsubsection{Performance under Different System Configurations}
To quantify the impact of system architecture, we evaluated transformer inference throughput under four memory/interconnect configurations: (1) using only host memory with a low-bandwidth PCIe link (2 GB/s, 4 lanes @ 4 Gbps), (2) host memory with a moderate PCIe bandwidth (8 GB/s, 8 lanes @ 8 Gbps), (3) host memory with a high-bandwidth interconnect (64 GB/s, 16 lanes @ 64 Gbps), and (4) a system with device-side memory (no host memory accesses for the model, effectively a NUMA node with HBM attached to the accelerator). These options range from a constrained host–accelerator link to an idealized on-card memory setup. We applied best-case optimizations from our microbenchmarks (e.g., 256B DMA transfers and appropriate memory controllers for each scenario) to ensure that each configuration was as efficient as possible given its constraints.

\begin{figure}[htbp]
    \centering
    \includegraphics[width=0.8\linewidth]{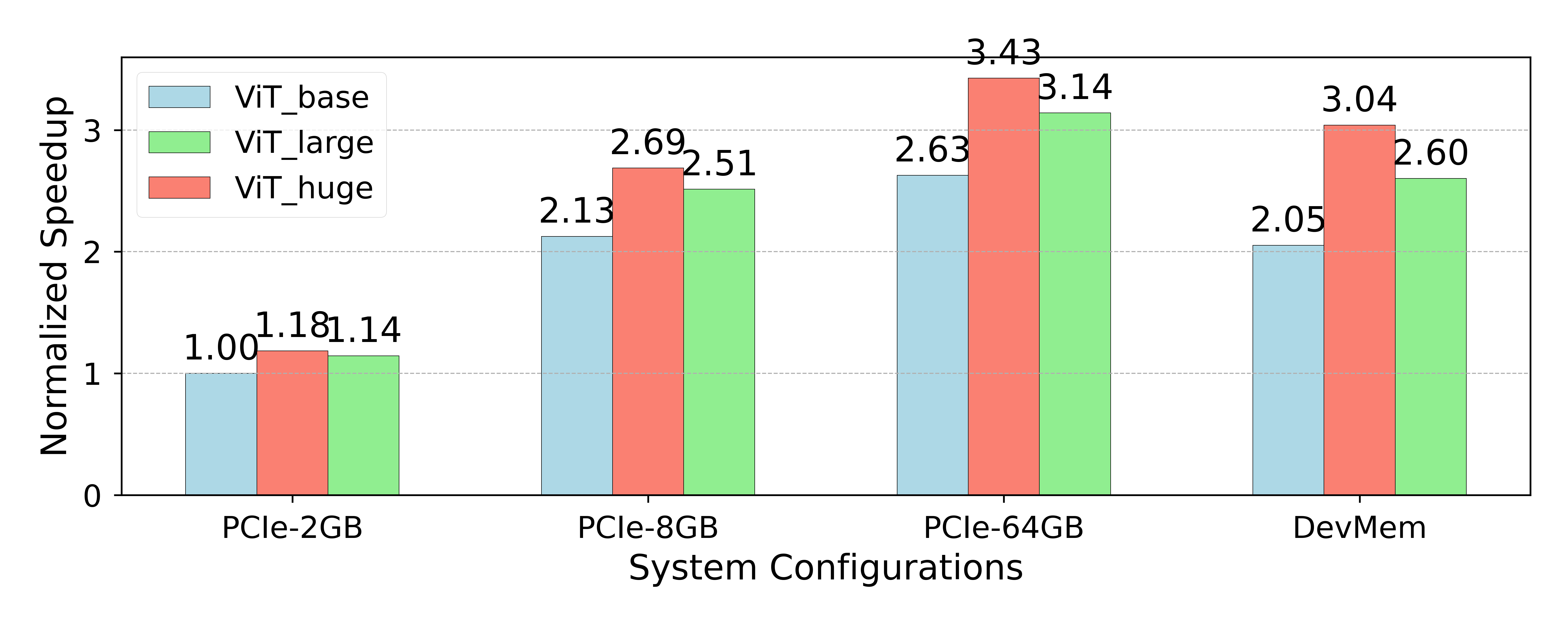}
    \vspace{-0.25cm}
    \caption{Performance comparison of memory locations and interconnects~\cite{Liu2025AcceSys}}
    \Description{Performance comparison of memory locations and interconnects}
    \label{fig:config_comparison}
    \vspace{-0.25cm}
\end{figure}

Figure~\ref{fig:config_comparison} summarizes the normalized inference throughput for three representative models (ViT-Base, ViT-Large, ViT-Huge) under these four configurations. Several clear trends emerge. First, increasing the PCIe bandwidth dramatically improves end-to-end performance when using host memory. Increasing the link from 2GB/s to 8GB/s and 64GB/s yields speedups of roughly 2.5× and 3×, respectively. For example, ViT-Huge runs about 3.4× faster with a 64GB/s interface compared to a 2GB/s link. This shows that a slow PCIe interface can bottleneck the entire transformer workload and that a high-bandwidth standard interconnect (e.g., a PCIe Gen5 link and beyond) can largely bridge the gap. Second, the on-card memory configuration (DevMem) achieves competitive performance, but does not vastly outperform the best host-memory scenario. In fact, in our results, a purely host-memory system with a 64GB/s PCIe link slightly outpaced the DevMem setup by roughly 5–15\% for the same models. For instance, on ViT-Huge the host-memory + 64GB/s PCIe configuration delivered ~1.13× the throughput of the HBM-based (DevMem) system. This outcome is initially surprising given the accelerator’s fast local HBM2 memory, but the explanation lies in system-level overheads for the portions of the workload that run on the host CPU. Simply adding on-card high-bandwidth memory does not guarantee a speedup if the rest of the system (software, data movement, workload distribution) is not co-optimized to take advantage of it. In summary, aggressively maximizing standard interconnect bandwidth can allow a host-memory design to approach or even exceed the performance of a device-memory design, offering a potential cost advantage (using commodity host memory) when the workload characteristics are favorable.

To complement the normalized throughput of Fig.~\ref{fig:config_comparison}, we report wall-clock, end-to-end latency (batch = 1) and the corresponding frame rate for ViT with $224\times224$ inputs. For \textbf{ViT-Large}, a constrained host link of 2~GB/s processes one image in 2.98~s ($\approx$0.34~FPS), improving to 1.108~s ($\approx$0.90~FPS) at 8~GB/s and 0.98~s ($\approx$1.02~FPS) at 64~GB/s; the DevMem (HBM) setup completes in 1.13~s ($\approx$0.89~FPS). For \textbf{ViT-Base}, which is 4.45$\times$ faster under the same conditions, the latencies are 669.66~ms ($\approx$1.49~FPS), 248.99~ms ($\approx$4.02~FPS), and 220.22~ms ($\approx$4.54~FPS) for 2, 8, and 64~GB/s host links, respectively, and 253.93~ms ($\approx$3.94~FPS) for DevMem. Increasing host-side bandwidth from 2~GB/s to 64~GB/s reduces per-image latency by \textbf{2000~ms} for ViT-Large and \textbf{449.44~ms} for ViT-Base, corresponding in both cases to an FPS gain of about \textbf{3.04}$\times$. These absolute numbers make the interconnect effect explicit and explain why a well-provisioned host-memory plus fast PCIe path can match or slightly exceed DevMem in end-to-end ViT throughput on our setup (Fig.~\ref{fig:config_comparison}).

\subsubsection{Analysis of GEMM vs. Non-GEMM Operations}
\label{subsec:gemmana}
\begin{figure}[htbp]
    \centering
    \includegraphics[width=0.8\linewidth]{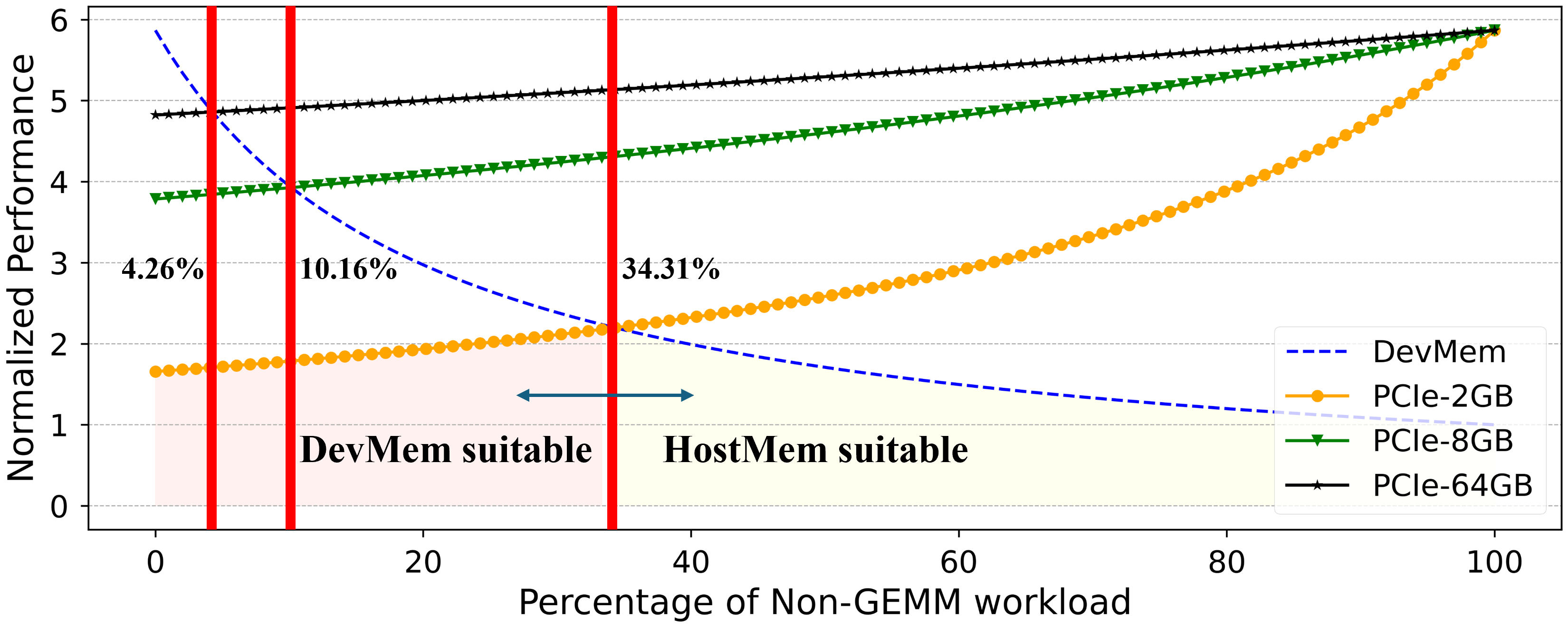}
    \vspace{-0.15cm}
    \caption{Overall Transformer Performance as a Function of Non-GEMM Workload Fraction for various PCIe Bandwidths vs DevMem~\cite{Liu2025AcceSys}}
    \Description{Overall Transformer Performance as a Function of Non-GEMM Workload Fraction}
    \label{fig:NonGEMMPer_comparison}
    \vspace{-0.25cm}
\end{figure}

Figure~\ref{fig:NonGEMMPer_comparison} examines overall accelerator performance as we vary the fraction of non-GEMM computations in the workload, comparing an all-DevMem system to host-memory systems with different PCIe bandwidths (performance is normalized to the DevMem case). The key trend is clear: as the non-GEMM portion of the workload grows, the advantage of keeping all data in device memory rapidly erodes. When the model is almost entirely matrix computations (non-GEMM <$\sim$5\%), the DevMem configuration yields the best performance – the accelerator handles almost all work locally and avoids PCIe transfer overhead. However, once even a modest share of the workload is non-GEMM (e.g., >$\sim$35\%), a fast host-memory system can catch up to and eventually surpass the DevMem baseline. This crossover occurs because in the DevMem scenario every CPU-side operation (the non-GEMM work) incurs a penalty to access data over PCIe. As the non-GEMM fraction increases, those penalties accumulate and decrease DevMem’s benefit.
By contrast, using host memory means that the CPU can perform non-GEMM tasks without off-chip delays, and only the GEMM portions incur PCIe transfer costs. At higher non-GEMM fractions, the time saved on CPU operations outweighs the extra time spent transferring matrices for GEMMs. In other words, for transformer models with a moderate or large amount of non-matrix computation, a well-tuned host-memory system – especially with a high-bandwidth PCIe interface – will eventually overtake an all-DevMem approach in total throughput. In fact, Figure~\ref{fig:NonGEMMPer_comparison} shows that with a PCIe link of the top end (64GB / s), the host memory configuration begins to outperform DevMem once the workload is no longer almost exclusively GEMM. Slower PCIe links require a larger non-GEMM share to flip this balance, but the overall pattern remains the same: the more non-GEMM work, the less DevMem’s initial lead matters.

MatrixFlow’s loosely-coupled design is well-suited to exploit this trade-off. Because the accelerator is not tied to a fixed on-card memory pool, it can flexibly use host memory and PCIe bandwidth to handle non-GEMM portions of the workload without stalling the compute units. The result is that for realistic transformer inference tasks (which include operations like attention softmax, normalization, etc., not just GEMMs), a balanced system with high-speed host memory access can outperform a device-memory-only setup. This insight underscores an important architectural point: Beyond a very small fraction of non-GEMM, simply keeping all data in accelerator local memory is suboptimal – leveraging a fast host interface for non-GEMM parts yields better overall performance.

\subsubsection{Comparison with Existing Accelerators}
We finally compare our co-designed MatrixFlow platform with other state-of-the-art accelerator solutions for transformer inference. Table~\ref{tab:transperformance} summarizes the performance (throughput or speed-up) of several systems on transformer models, including a multi-core CPU baseline, a prior loosely-coupled accelerator simulation (SMAUG), and a tightly-coupled systolic array accelerator (TiC-SAT), alongside our MatrixFlow results. For fair comparison, all accelerators are evaluated within a full-system simulation environment (gem5-based), and performance is normalized to a single-core CPU baseline. The many-core CPU (OpenMP) system achieves about 24–26× speed-up over one core on these transformer workloads, which underscores strong scaling, but also the limitations of general-purpose processors (performance plateaus for larger models due to memory bottlenecks and limited vectorization). The SMAUG platform (a gem5-based simulator with an integrated accelerator model) reports up to 88× speed-up over the single-core baseline. This loosely-coupled design significantly improved over the CPU by offloading to a custom unit, but it used Float16 precision and relatively limited off-chip bandwidth. The tightly-coupled TiC-SAT accelerator (integrated as an ISA extension using int8 arithmetic) performs even better, with speed-ups ranging from ~58× on smaller transformers to ~89× on large ones. TiC-SAT benefits from minimal communication overhead (the accelerator is in-core) and aggressive data reuse optimizations, showing good scaling on larger models. 

\begin{table}[htbp]
  \centering
  \caption{Transformer Performance Comparison~\cite{Liu2025}}
  \vspace{-0.1cm}
    \begin{tabular}{lcccccc}
      \toprule
      \textbf{Configuration} & \multicolumn{3}{c}{\textbf{BERT Models}} & \multicolumn{3}{c}{\textbf{ViT Models}} \\
      \cmidrule(lr){2-4} \cmidrule(lr){5-7}
      & \textbf{Medium} & \textbf{Base} & \textbf{Large} & \textbf{Base} & \textbf{Large} & \textbf{Huge} \\
      \midrule
      Single-thread   & 1    & 1    & 1    & 1    & 1    & 1 \\
      Multi-threaded  & 23.7 & 24.3 & 25.6 & 23.7 & 24.3 & 25.6 \\
      SMAUG~\cite{Xi2020}           & 88   & -    & -    & -    & -    & - \\
      TIC-SAT~\cite{Amirshahi2023}         & 58.3 & 69.3 & 89.5 & 69.4 & 82.5 & 82.7 \\
      MatrixFlow      & 453.9& 633.7& 698.2& 327.9& 392.0& 427.6\\
      \bottomrule
    \end{tabular}
  \label{tab:transperformance}
  \vspace{-0.15cm}
\end{table}

Against this backdrop, MatrixFlow achieves the highest performance. Our design reaches up to ~698× single-core speed-up on BERT-Large, versus ~89× for TiC-SAT on the same model. That is an 8× improvement over the state-of-the-art tightly-coupled accelerator. Similarly, against the loosely-coupled SMAUG, MatrixFlow is about 5× faster (e.g., 453× vs 88× on a medium transformer). These gains stem from the holistic hardware/software co-design: MatrixFlow uses int32 precision to avoid any accuracy loss (unlike float16 in SMAUG) yet still excels due to its optimized dataflow, efficient DMA/PCIe usage, and the elimination of unnecessary memory buffering. In fact, MatrixFlow’s design uses only three small 4KB on-chip buffers (one for each operand and for results) – drastically smaller than other accelerators that dedicate tens or hundreds of KB to SRAM scratchpads. This lean buffering is possible because the accelerator pipelines data directly from memory in a streaming fashion, enabled by our high-throughput PCIe and memory subsystem co-design. The result is a system that outperforms both loosely and tightly coupled prior accelerators, while maintaining flexibility (MatrixFlow is loosely coupled and does not require custom ISA changes). It is worth noting that MatrixFlow also handily beats a many-core CPU baseline; in our experiments, it was up to 22× faster than a 64-thread CPU system on end-to-end transformer inference. Overall, these comparisons validate that our system-level approach – focusing not just on the accelerator array, but also on data orchestration, standard interconnect integration, and software mapping – yields state-of-the-art performance for transformer workloads.

\section{Conclusion}
\label{sec:conclusion}
This work has demonstrated that a holistic hardware/software co-design strategy can dramatically improve transformer inference performance. By co-optimizing a compact 16×16 systolic-array accelerator (MatrixFlow) together with its full-system integration in gem5-AcceSys, we achieved a near-streaming execution of transformer layers with minimal on-chip buffering and maximized data throughput. The result is a lean accelerator system that delivers up to 22× faster end-to-end inference latency than a CPU-only baseline and outperforms state-of-the-art accelerators by 5×–8×, all while using standard system interfaces and without requiring any specialized host processor modifications. These results underscore the potency of co-design, shifting data orchestration into the system (leveraging conventional PCIe, DMA, and memory hierarchy resources) allows the accelerator to avoid idle stalls and fully utilize its compute capacity, and provide a novel example on the architecture exploration with the combination of lightweight accelerator and data streaming accelerator.

Beyond raw performance gains, our co-design approach yields broader system-level insights. Full-system analysis revealed that memory bandwidth, rather than latency, is the dominant factor in sustaining high-transmission throughput. This finding suggests that future AI accelerators must be architected with balanced system integration in mind, ensuring ample memory bandwidth and efficient interconnects to keep the compute engines fed. The Gem5-AcceSys framework proved instrumental in exposing such bottlenecks, showing that aligning accelerator microarchitecture with realistic integration constraints can unlock exceptional performance without any custom host support. In summary, the joint development of MatrixFlow and Gem5-AcceSys redefines the performance limits of transformer inference and provides a practical blueprint for next-generation AI accelerators, illustrating the effectiveness of a balanced co-design paradigm in modern AI systems.
\vspace{-0.25cm}
\begin{acks}
This work was supported in part by the Swiss State Secretariat for Education, Research, and Innovation (SERI) through the SwissChips research project, and also by Intel as part of the Intel Center for Heterogeneous Integrated Platforms (HIP).
\end{acks}
\vspace{-0.25cm}
\bibliographystyle{ACM-Reference-Format}
\bibliography{acmart}

\end{document}